\documentclass[preprint,12pt,authoryear]{elsarticle}

\usepackage{amssymb}
\usepackage{amsmath}
\usepackage{float}

\usepackage{mathtools}
\usepackage{graphicx}
\usepackage[dvipsnames]{xcolor}
\usepackage{tikz}
\usepackage{pgfplots}
\definecolor{colorA}{rgb}{0.12156862745098039,0.4666666666666667,0.7058823529411765}
\definecolor{colorB}{rgb}{1.0,0.4980392156862745,0.054901960784313725}
\definecolor{colorC}{rgb}{0.17254901960784313,0.6274509803921569,0.17254901960784313}
\definecolor{colorD}{rgb}{0.8392156862745098,0.15294117647058825,0.1568627450980392}
\definecolor{colorE}{rgb}{0.5803921568627451,0.403921568627451,0.7411764705882353}
\definecolor{colorF}{rgb}{0.5490196078431373,0.33725490196078434,0.29411764705882354}
\definecolor{colorG}{rgb}{0.5019607843137255,0.5019607843137255,0.5019607843137255}
\definecolor{colorH}{rgb}{0.0,0.0,0.0}

\journal{International Journal of Heat and Fluid Flow}

\usepackage[normalem]{ulem}
\usepackage[dvipsnames]{xcolor}
\definecolor{brown}{rgb}{0.5,0.1,0.05}

\begin{document}

\begin{frontmatter}


\title{Predicting the wall-shear stress and wall pressure through convolutional neural networks}


\affiliation[insa]{organization={FLOW, Engineering Mechanics, KTH Royal Institute of Technology},city={SE-100 44  Stockholm}, country={Sweden}}
\affiliation[insb]{organization={Swedish e-Science Research Centre (SeRC)},city={SE-100 44  Stockholm},country={Sweden}}
\affiliation[insd]{organization={Division of Robotics, Perception, and Learning, KTH Royal Institute of Technology},city={SE-100 44  Stockholm},country={Sweden}}

\author[insa,insb]{A. G. Balasubramanian\corref{cor1}}
\ead{argb@kth.se}
\cortext[cor1]{Corresponding authors}
\author[insa,insb]{L. Guastoni}
\author[insa,insb]{P. Schlatter}
\author[insd,insb]{H. Azizpour}
\author[insa,insb]{R. Vinuesa\corref{cor1}}
\ead{rvinuesa@mech.kth.se}

\begin{abstract}
The objective of this study is to assess the capability of convolution-based neural networks to predict the wall quantities in a turbulent open channel flow, starting from measurements within the flow. Gradually approaching the wall, the first tests are performed by training a fully-convolutional network (FCN) to predict the two-dimensional velocity-fluctuation fields at the inner-scaled wall-normal location $y^{+}_{\rm target}$, using the sampled velocity fluctuations in wall-parallel planes located farther from the wall, at $y^{+}_{\rm input}$. 
The predictions from the FCN are compared against the predictions from a proposed R-Net architecture as a part of the network investigation study. 
Since the R-Net model is found to perform better than the FCN model, the former architecture is optimized to predict the two-dimensional streamwise and spanwise wall-shear-stress components and the wall pressure from the sampled velocity-fluctuation fields farther from the wall.
The data for training and testing is obtained from direct numerical simulation (DNS) of open channel flow at friction Reynolds numbers $Re_{\tau} = 180$ and $550$. The turbulent velocity-fluctuation fields are sampled at various inner-scaled wall-normal locations, {\it i.e.} $y^{+} = \{15, 30, 50, 100, 150\}$, along with the wall-shear stress and the wall pressure. At $Re_{\tau}=550$, both FCN and R-Net can take advantage of the self-similarity in the logarithmic region of the flow and predict the velocity-fluctuation fields at $y^{+} = 50$ using the velocity-fluctuation fields at $y^{+} = 100$ as input with about 10\% error in prediction of streamwise-fluctuations intensity. Further, the network model trained in this work is also able to predict the wall-shear-stress and wall-pressure fields using the velocity-fluctuation fields at $y^+  = 50$ with around 10\% error in the intensity of the corresponding fluctuations at both $Re_{\tau} = 180$ and $550$.
These results are an encouraging starting point to develop neural-network-based approaches for modelling turbulence near the wall in numerical simulations, especially large-eddy simulations (LESs).
\end{abstract}




\end{frontmatter}


\section{Introduction}
\label{sec:section1_introduction}


The direct numerical simulation (DNS) of turbulent flows is computationally challenging owing to the cost arising from the resolution requirements to simulate all the length and time scales of fluid motion. Turbulent flows in various engineering applications and practical flows of interest are characterized by a very high Reynolds number and it becomes computationally unfeasible to simulate the wide range of scales associated with these flows. For this reason, simpler canonical flow cases like the channel flows~\citep{hoyas} and turbulent boundary layers~\citep{sillero,pozuelo} are widely studied using DNSs at relatively low Reynolds number to understand the physics of flow and thereby to model them. At higher Reynolds numbers, several investigations have been performed using high-resolution large-eddy simulations~\citep{schlatter,bobke,li}. Apart from canonical flows, LES has also been widely employed for different practical flows of interest like urban flows~\citep{atzori} and the flow around an airfoil~\citep{vinuesa_2018,tamaki}. 

Large eddy simulations address the issue of computational cost by filtering out the smallest turbulent scales and modelling the eddies that are smaller than the filter size. However, it should be noted that despite modelling the effects of the smallest scales, LESs can become computationally expensive at high Reynolds number and especially in wall-bounded flows. This is due to the resolution requirements for resolving the dynamically important flow structures in the viscous and the logarithmic layers, which scale as $\mathcal{O} (Re_{\tau}^2)$~\citep{larsson}, where the friction Reynolds number $Re_{\tau}$ is defined in terms of reference length $h$ and friction velocity $u_{\tau} = \sqrt{\tau_w/\rho}$ where $\tau_w$ is the wall-shear stress and $\rho$ is the fluid density. In wall-bounded turbulent flows, the size of the scales that characterize the near-wall region is very small. Hence, a wall-resolved LES can have a significant cost with increasing Reynolds number~\citep{choi} and approaches to that of direct numerical simulations which prevents its use in high Reynolds number wall-bounded flows that occur in engineering applications~\citep{park}.. This problem is addressed by introducing wall models, which allow a manageable resolution while taking into account the characteristic of the flow related to wall presence. 

Several methods have been proposed to model the near-wall region and simulate complex flows at higher Reynolds numbers. Such methods mainly focused on wall-stress-modelling in addition to the hybrid LES/Reynolds-averaged Navier Stokes (RANS) approach.
These wall models aim to reproduce the most important features of the inner-layer dynamics like (\textit{e.g.} streaks and streamwise vortices), without integrating the Navier--Stokes equations in that flow region. Conventionally, the wall models were either based on physics or mathematical arguments~\citep{larsson}. Focusing our attention towards wall-shear stress modelling, the algebraic equilibrium wall model is the most commonly used method to specify wall-shear stress owing to its simplicity, which enforces the law of the wall both locally and instantaneously~\citep{schuman} based on the idea that turbulence close to the wall reaches a state of equilibrium where the production and dissipation are balanced. However, the equilibrium assumption may not hold for flows with adverse pressure gradients or for cases with strong non-equilibrium effects. In particular, the integral wall model aims to address the non-equilirbium effects by solving a vertically integrated momentum equation which adds to the equilibrium logarithmic velocity profile~\citep{yang}. Other studies like~\cite{wang} and \cite{kawai2012} have also explored the non-equilibrium wall models which involve solving RANS equations. On the other hand, zonal model solves the thin-boundary-layer equation on a set of refined mesh near the wall~\citep{park,larsson}. The wall-shear stress fluctuations modelling~\citep{inoue} and dynamic slip wall models~\citep{bose2014,bae2018} are few other approaches to obtain the required wall-shear stress. A classification of wall models with the corresponding assumptions and error sources is presented in~\cite{piomelli}.
A recent review of wall models can be found in~\cite{larsson} and~\cite{bose}.

One alternative approach was proposed by~\cite{mizuno}: by taking advantage of the self-similarity hypothesis, according to which the eddy sizes scale linearly in the logarithmic region at high Reynolds number, an \emph{off-wall} boundary condition was defined. This approach moves the boundary of the computational domain away from the wall, removing the need to simulate the small scales associated with it. In their work, the velocity field at the off-wall boundary location is substituted by a re-scaled and a shifted copy of an interior reference plane farther from the wall. This allows the logarithmic region to be simulated without considering the near-wall dynamics. In this work, we investigate the possibility to train a neural-network model that predicts the flow close to the wall based on information farther from the wall, with the future possibility to apply it as an off-wall boundary condition.

Recently, machine-learning-based approaches have been increasingly used in the study of turbulent flows for a variety of tasks related to flow prediction~\citep{duraisamy,brunton_2020,guastoni_2021,guastoni_2022}, prediction of temporal dynamics~\citep{srinivasan,eivazi,borrelli}, extraction of flow patterns~\citep{jimenez_2018,eivazi_2022,martinez}, generation of inflow conditions \citep{fukami,yousif} or flow control~\citep{rabault,vinuesa_22,guastoni_2023} to name a few. In addition, neural network models have started offering new interesting opportunities to formulate efficient data-driven wall models, as highlighted in~\cite{vinuesa_2022}.~\cite{yang2019} utilized physics informed neural network (PINN) to propose a wall model, where the knowledge of mean flow scaling was incorporated in the feed-forward neural network to predict the LES-grid filtered wall-shear stress from the velocity at a distance from the wall. Their predictive model was shown to capture the law of the wall at different Reynolds number which was not present in the training dataset.~\cite{zhou2021} trained a feed-forward neural network to predict the wall-shear stress as a function of velocity and pressure gradients at a off-wall location in a periodic hill flow. Although the \emph{a priori} tests showed good results, \emph{a posteriori} tests did not provide accurate results. Whereas towards defining an off-wall boundary condition, the works like~\cite{moriya} and~\cite{bae} have exhibited promising potential of deep-learning and reinforcement learning based approaches in predicting the flow close to the wall. A survey of machine-learning-based wall models for large-eddy simulations can be found in~\cite{vardot}. The advantage of such machine-learning-based models is that once the model is trained, the evaluation of the neural network is computationally cheap and they can provide a valid alternative to the models that are currently employed within numerical simulations. 

In the present work, we first employ a fully convolutional network to predict the near-wall velocity-fluctuation fields using the velocity-fluctuation fields farther from the wall, corresponding to $Re_{\tau} = 180$. In addition, we explore the capability of a fully-convolutional network with residual skip-connections (denoted as R-Net) in predicting the near-wall velocity fluctuation fields. Then, the described network models (FCN and R-Net) are used to predict the velocity-fluctuation fields closer to the wall that is present in the self-similar region at $Re_{\tau} = 550$. A brief comparison of the results obtained from FCN and R-Net is performed, in order to provide an insight to designing an efficient network model in predicting wall quantities. Finally, the R-Net architecture model is used to predict the wall-shear stress and wall pressure using the velocity-fluctuation fields away from the wall.

The manuscript is structured as follows. An overview of the methodology employed in this study is provided in~$\S$\ref{sec:section2_methodology}: the architecture of the neural-network models are introduced in~$\S$\ref{subsec:Network_architecture} and the dataset utilized to train the network models is described in~$\S$\ref{subsec:Dataset}. The metrics used to compare the performance of the different networks are reported in~$\S$\ref{subsec:performance}. The results are discussed in~$\S$\ref{sec:section3_results},  with~$\S$\ref{subsec:Inner_predictions}--$\S$\ref{subsec:wall_predictions} containing the results corresponding to inner predictions, predictions in the self-similar region, network comparison and wall predictions, respectively. Finally, the conclusions are provided in~$\S$\ref{sec:conclusion}.

\section{Methodology}
\label{sec:section2_methodology}

Taking inspiration from the data-driven approach proposed by~\cite{guastoni_2021} for \textit{non-intrusive} sensing, where the turbulent velocity field at a given wall-normal distance is predicted by using quantities measured at the wall as inputs, we aim to explore the capability of convolutional networks to solve the opposite problem, where the inputs are located at a certain wall-normal distance with the objective to predict the wall-shear quantities and wall-pressure. We denote the prediction of wall quantities using the fields farther from the wall as \emph{wall predictions}.
As a primary step towards understanding the capability of convolutional networks in predicting the wall quantities, first we consider a simpler problem which is to predict the velocity-fluctuation fields closer to the wall at $y^+_{\rm target}$ using the velocity-fluctuation fields farther from the wall at $y^+_{\rm input}$ as inputs. Note that `+' denotes inner scaling, {\it i.e.} scaling with the friction velocity $u_{\tau}$ and the fluid kinematic viscosity $\nu$. Such predictions where $y^+_{\rm train} > y^+_{\rm target}$ are designated as \emph{inner predictions} whereas, the cases in which $y^+_{\rm train} < y^+_{\rm target}$ are denoted as \emph{outer predictions}. We test the performance of the networks for inner predictions by analyzing the mean-squared error in the instantaneous predictions and in the turbulent statistics as a function of the distance between the input and target velocity plane, $\Delta y^+ \left(||y^+_{\rm input} - y^+_{\rm target} ||\right)$. This parametric study is performed at a lower $Re_{\tau}$ of $180$, however a selected case is considered at a higher Reynolds number, where self-similarity can be exploited by the prediction model.

The assumption of self-similarity in a channel of height $2h$ for the two wall-normal planes at $y/h = 0.2$ and $y/h = 0.1$ (where the length scales are proportional to $y$) at friction Reynolds number $Re_{\tau} \approx 1000$ was the starting point to design an off-wall boundary condition in the model by~\cite{mizuno}. In the present work, we perform predictions in the self-similar region at a comparatively lower Reynolds number of $Re_{\tau} = 550$, with the wall-normal planes corresponding to $y/h = 0.2$ and $y/h = 0.1$ in scaled outer units. Such wall-normal planes correspond to $y^{+} = 100$ and $y^{+} = 50$, in inner-scaling respectively.

The predictions from fully-convolutional networks (FCNs) are compared with the ones from R-Net architecture models.
The R-Net architecture is then trained to predict the wall-shear and wall pressure quantities using the velocity fluctuation fields as inputs at different wall-normal distances. The prediction of wall quantities using R-Net are summarized for the two-different Reynolds number considered in this study.

\subsection{Network architecture}
\label{subsec:Network_architecture}

\begin{figure}[H]
\begin{center}
\includegraphics*[width=1.0\linewidth]{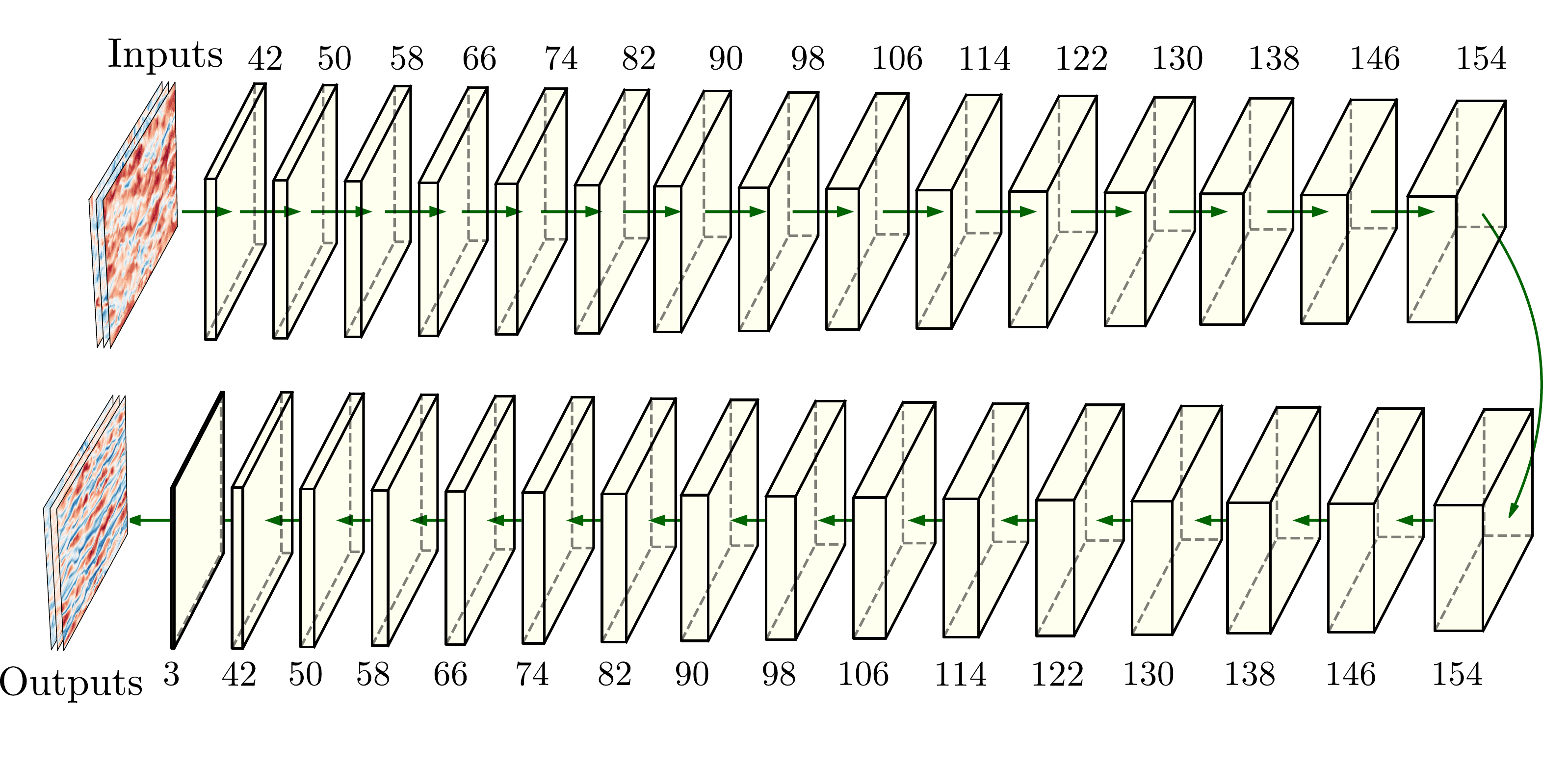}
\includegraphics*[width=1.0\linewidth]{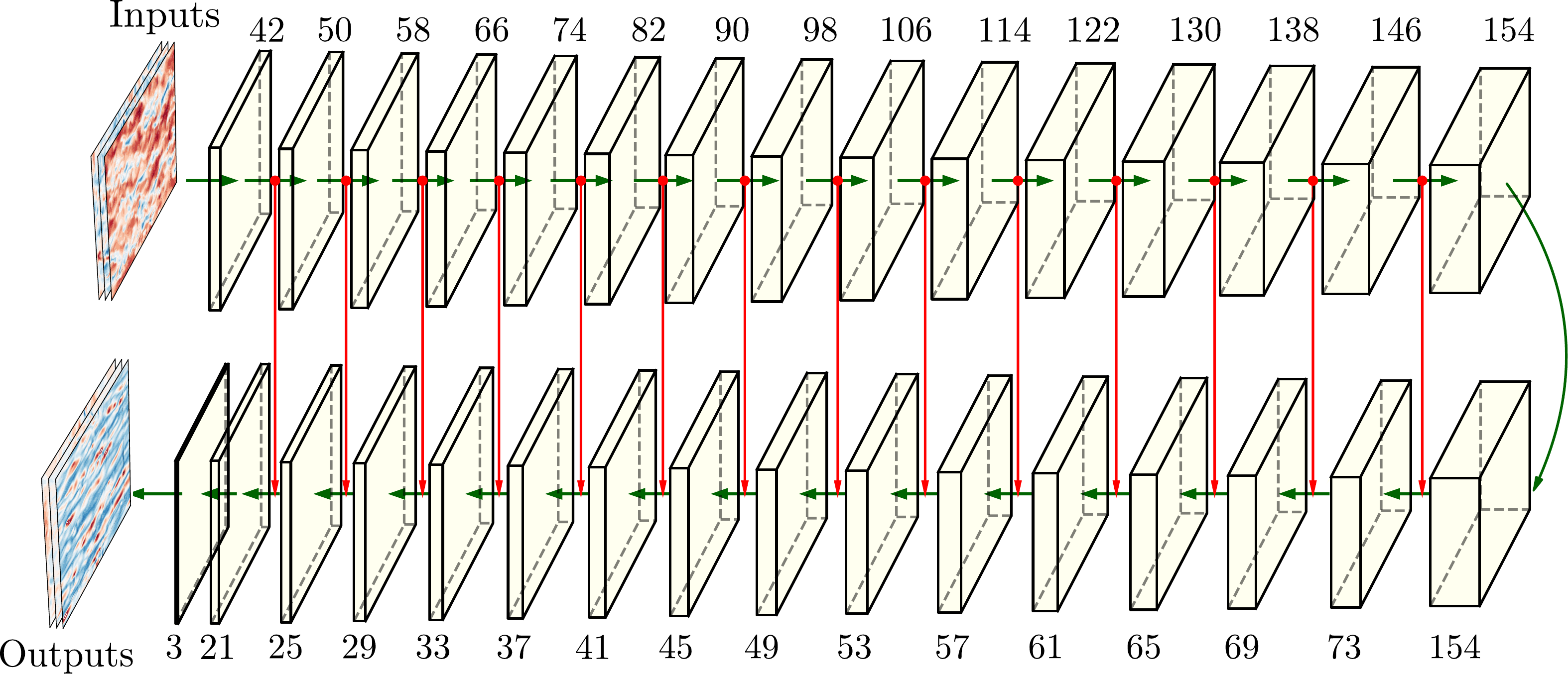}
\caption{\label{fig:cnn_arch} Schematic representation of the (top) FCN architecture and (bottom) R-Net architecture. The number of kernels applied to each of the layers is indicated in the figure. Here, the inputs correspond to the velocity-fluctuation fields at $y^+ =  50$ and the outputs in FCN are the velocity-fluctuation fields at $y^+=15$. In R-Net, it corresponds to the wall-shear stress components and the wall pressure. The skip-connections from the upstream part of the R-Net architecture to the downstream part of the network are indicated in red.}   
\end{center}
\end{figure}

A FCN similar to the one proposed by~\cite{guastoni_2022} is used in this study which adopts convolution operations in each layer~\citep{lecun} as given by:
\begin{equation}
    F_{i,j} = \sum_{m}\sum_{n} I_ {i-m,j-n}K_{m,n}\,,
\end{equation}
 where $\mathbf{I} \in \mathbb{R}^{d_{1}\times d_{2}}$ is the input, $\mathbf{K} \in \mathbb{R}^{k_{1} \times k_{2}}$ is the kernel (or filter) containing the parameters to be learned and $\mathbf{F}$ is the resulting feature map. The convolutional kernels are defined by their width and height and in this study, kernels of the size $3\times 3$ is used. A convolutional layer is defined by the number of kernels and the size of the kernel, which contain the learnable parameters. The transformed output is called \textit{feature map}. Multiple feature maps are usually stacked and followed by an element-wise activation function to make each layer a non-linear transformation. The non-linearity in the activation of layers is introduced by the rectified linear unit (ReLU) activation function. The output of the activation function is fed into another convolutional layer, which allows the next layer to combine the features individually identified in each map. This enables the prediction of larger and more complex features for progressively deeper networks. Batch normalization~\citep{batch} is performed after each convolutional layer (except for the last one). Since the ReLU activation function filters out the negative values, a modified ReLU function is used just before the output layer with a threshold of $-1$ (for prediction of velocity fluctuations) or $-25$ (for prediction of fluctuation wall quantities). The schematic of a fully convolutional network used in this study is shown in figure~\ref{fig:cnn_arch}. The FCN employed in this study has 31 hidden layers with a total of 2,902,791 trainable parameters. A fully-convolutional network with skip-connections~\citep{resnet} between hidden layers ${i}$ and ${N-i-1}$ (where $N$ indicates the total number of hidden layers in the network) is also investigated in this study and is denoted as R-Net. Note that the skip-connections resemble that of the ones employed in a typical U-Net architecture~\citep{unett}. R-Net is an intermediary step between FCN and U-Net which does not employ up-sampling as present in a U-Net. The proposed R-Net architecture is shown in figure~\ref{fig:cnn_arch}. The feature maps in the upstream part of the network are of different size compared to the ones in the downstream part of the network. Hence while performing skip connections, the feature maps from the upstream part of the network is cropped to the desired shape and concatenated to the feature maps in the downstream part of the network. The R-Net architecture consists of a total of 31 hidden layers with 2,568,681 trainable parameters.

\subsection{Dataset}
\label{subsec:Dataset}
The convolutional networks are trained to predict two-dimensional velocity-fluctuation fields at the given wall-normal location $y^{+}_{\rm target}$, using velocity-fluctuation fields farther from the wall, at $y^{+}_{\rm input}$, as inputs (where $y^{+}_{\rm input} > y^{+}_{\rm target}$).
The velocity-fluctuations fields are sampled from a direct numerical simulation of an open channel flow, at friction Reynolds numbers $Re_{\tau} = 180$ and $Re_{\tau} = 550$, performed with the pseudo-spectral solver SIMSON~\citep{simson}. The reference length $h$ for this flow case is the open-channel height and the sampled wall-parallel fields have size $L_x \times L_z =  4\pi h\times 2\pi h$. Periodic boundary conditions are imposed in the wall-parallel directions.
The turbulent flow fields are sampled at wall-normal locations, $y^{+}=15,30,50,100$ and $150$. The resolution of the sampled fields is $(N_{x},N_{z}) = (192,192)$ for $Re_{\tau} = 180$ and $(512, 512)$ for $Re_{\tau} = 550$. Further details on the generation of dataset are available in~\cite{guastoni_2021}.

The velocity fields are stored with a constant sampling period of $\Delta t_{s}^{+} = 5.08$ for $Re_{\tau} = 180$ and at $\Delta t_{s}^{+} = 1.49$ for $Re_{\tau} = 550$. The networks are trained using 50,400 instantaneous fields for $Re_{\tau} = 180$ and with 19,920 fields for $Re_{\tau} = 550$, split into training and validation sets, with a ratio of 4 to 1. The input velocity fluctuation fields to the network are padded in the wall-parallel directions using a periodic padding operation, to enforce periodicity in the obtained predictions. In the padding operation, a total of 64 points (32 points on either side) are added along the boundary of the fields. Thereby the size of the input fields is $\approx77\%$ larger than that of the outputs for the fields at $Re_\tau = 180$ whereas, it is $\approx 26\%$ for the fields at $Re_\tau = 550$. Since the receptive field of the networks considered here is $63\times63$, the size of the outputs from the network is slightly larger than the sampled DNS flow fields and hence a cropping operation is performed to obtain the fields matching the size of the DNS fields. 

The input fluctuation fields to the network are scaled such that they have a similar magnitude as given by:
\begin{equation}
    \label{eqn_scaling}
    \hat{u} = u\,,\;\; \hat{v} = v\frac{u_{\rm RMS}}{v_{\rm RMS}}\,,\;\; \hat{w} = w\frac{u_{\rm RMS}}{w_{\rm RMS}}\,,
\end{equation}
where RMS refers to root-mean-squared quantities. The same scaling is used when the network outputs are velocity fields at a different wall-normal location.
For the prediction of fluctuations of the wall quantities, the outputs are normalized by the respective RMS quantities as defined by:
\begin{equation}
    \overline{\frac{\partial u}{\partial y}} = \frac{\partial u / \partial y}{\partial u/\partial y_{\rm RMS}}\,,\;\; \overline{\frac{\partial w}{\partial y}} = \frac{\partial w / \partial y}{\partial w/\partial y_{\rm RMS}}\,,\;\; \overline{p} = \frac{p}{p_{\rm RMS}}\,.
\end{equation}
In the remainder of the paper $\overline{\cdot}$ is dropped for clarity when referring to normalized quantities.

\subsection{Network performance measures}
\label{subsec:performance}

The mean-squared error (MSE) between the instantaneous DNS fields $(u_{\mathrm{DNS}})$ and the predictions ($u_{\mathrm{FCN}}$ denotes prediction from FCN; $u_{\mathrm{RN}}$ indicates prediction from R-Net) is used as the loss function for training the networks introduced in this study. It can be written as:

\begin{align}
\label{eqn_1}
\begin{split}
    \mathcal{L}(u_{\mathrm{FCN}};u_{\mathrm{DNS}}) &=
\frac{1}{N_{x}N_{z}} \sum_{i=1}^{N_{x}} \sum_{j=1}^{N_{z}} | u_{\mathrm{FCN}}(i,j) - u_{\mathrm{DNS}}(i,j) |^{2}\,, \\
\mathcal{L}(u_{\mathrm{RN}};u_{\mathrm{DNS}}) &=
\frac{1}{N_{x}N_{z}} \sum_{i=1}^{N_{x}} \sum_{j=1}^{N_{z}} | u_{\mathrm{RN}}(i,j) - u_{\mathrm{DNS}}(i,j) |^{2}\,.
\end{split}
\end{align}

The performance of the network is evaluated based on the quality of the instantaneous predictions in the test dataset, whose samples are obtained from simulations initialized with different random seeds, in order to ensure that the training and test datasets do not exhibit unwanted correlations. A total of 8,880 samples and 3,320 samples are used to obtain the converged statistics for $Re_{\tau} = 180$ and $Re_{\tau} = 550$, respectively. The sampling period for the test dataset is slightly different from the one used in the training dataset. We consider $\Delta t_s^+ = 1.69$ for $Re_{\tau} = 180$ and $\Delta t_s^+ = 1.49$ for $Re_{\tau} = 550$. It should be noted that the quality of instantaneous predictions improves when the network is trained with less-correlated samples (\textit{i.e. higher $\Delta t_{s}^{+}$}), provided that the network has sufficient capacity~\citep{guastoni_2021}. 

The predictions are also evaluated from the statistical point of view, namely considering the error in RMS quantities of velocity fluctuations:
\begin{align}
\label{eqn_2}
\begin{split}
E_{\mathrm{RMS; FCN}} (u) &= \frac{|u_{\mathrm{RMS,FCN}}-u_{\mathrm{RMS,DNS}}|}{u_{\mathrm{RMS,DNS}}}\,,\\
E_{\mathrm{RMS; RN}} (u) &= \frac{|u_{\mathrm{RMS,RN}}-u_{\mathrm{RMS,DNS}}|}{u_{\mathrm{RMS,DNS}}}\,,
\end{split}
\end{align}
and the instantaneous correlation coefficient between the predicted and the DNS fields:
\begin{align}
\label{eqn_3}
\begin{split}
R_{\mathrm{FCN; DNS}} (u) &= \frac{\left<u_{\mathrm{FCN}} u_{\mathrm{DNS}}\right>_{x,z,t}}{u_{\mathrm{RMS,FCN}} u_{\mathrm{RMS,DNS}}}\,,\\
R_{\mathrm{RN; DNS}} (u) &= \frac{\left<u_{\mathrm{RN}} u_{\mathrm{DNS}}\right>_{x,z,t}}{u_{\mathrm{RMS,RN}} u_{\mathrm{RMS,DNS}}}\,,
\end{split}
\end{align}
with $\left<\cdot\right>$ corresponding to the average in space or time, depending on the subscript. Note that the argument $u$ in equations~(\ref{eqn_2}) and~(\ref{eqn_3}) refers to the corresponding measures for streamwise velocity component. For wall predictions as discussed in~$\S$\ref{subsec:wall_predictions}, the argument is either $u_y$ or $w_y$ or $p$, which correspond to streamwise wall-shear stress, spanwise wall-shear stress and wall pressure, respectively. The sub-script `${\rm pred}$' is also used in the text to indicate the predicted quantity from the network. Finally, a comparison of the pre-multiplied two-dimensional power-spectral density is performed to verify the distribution of energy in different length scales. The two-dimensional (2D) pre-multiplied power-spectral density (PSD) $k_z k_x \phi_{ij} \left(\lambda_x^+,\lambda_z^+\right)$ is obtained in the streamwise and spanwise directions for the target DNS fields in the test dataset and is compared against the spectra obtained with the corresponding predictions from the network. Note that here $k_x$, $k_z$ denote wavenumbers in the streamwise and spanwise directions, with the corresponding wavelengths in inner-scaled units denoted as $\lambda_x^+$, $\lambda_z^+$. $\phi_{ij}$ is the
power-spectral density defined for the particular quantity $ij$.


\section{Results}
\label{sec:section3_results}
\subsection{Inner predictions}
\label{subsec:Inner_predictions}


In order to understand how the MSE is affected by the separation distance, a total of 10 predictions are performed with the velocity fields obtained at various wall-normal locations from the DNS simulation and we refer to them as \textit{inner} predictions, with different separation distances defined as: $\Delta y^{+} = y^{+}_{\rm input} - y^{+}_{\rm target}$. The input velocity-fluctuation field closest to the wall is located at $y^{+} = 30$ and the corresponding target field at $y^{+} = 15$. The training of the FCN and R-Net used for inner predictions consisted of 50 epochs, where an epoch identifies a complete pass through the data in the training dataset.

\begin{figure}
  \begin{minipage}[b]{0.5\textwidth}
    \includegraphics[width=\textwidth]{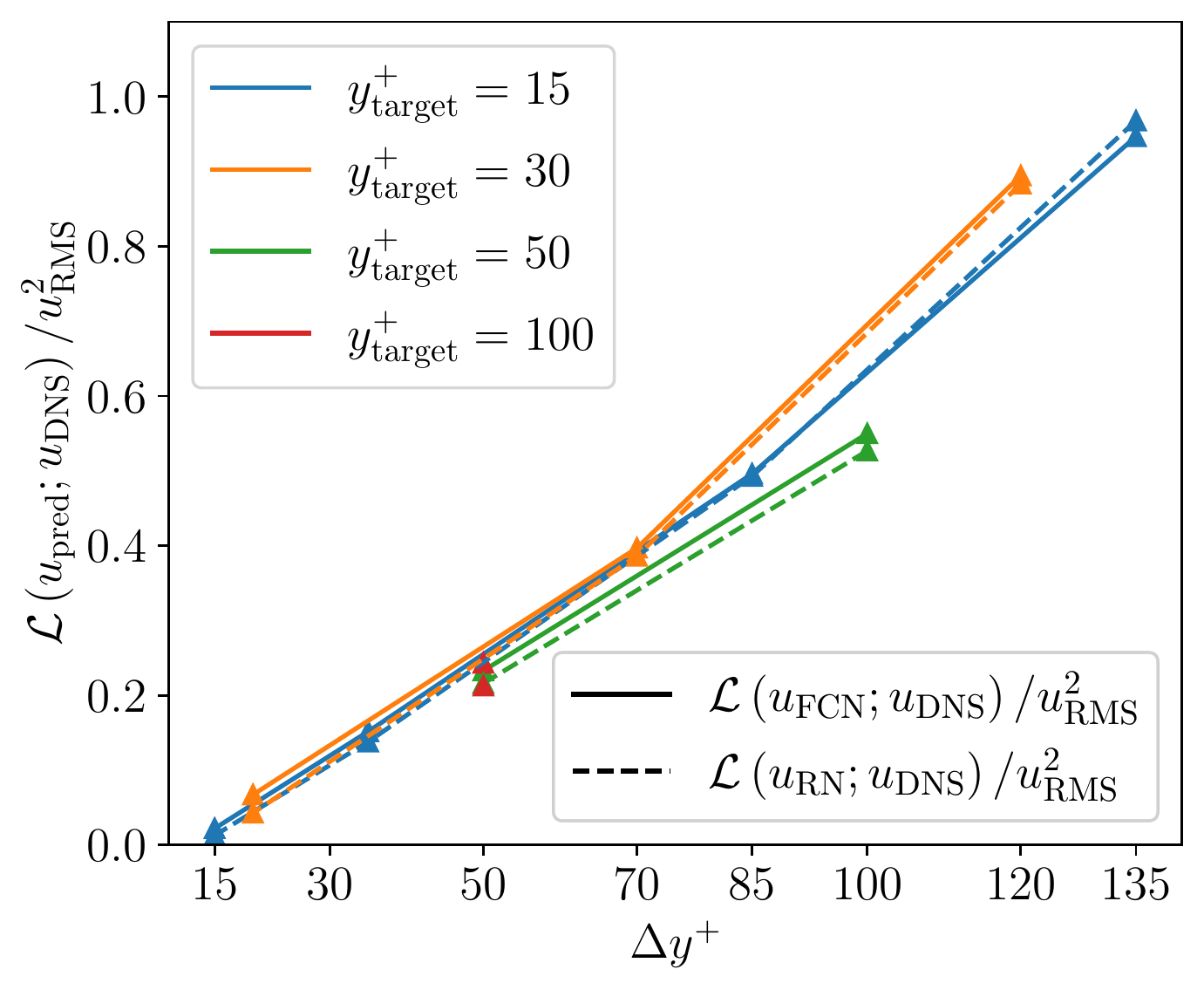}
  \end{minipage}
  \hfill
  \begin{minipage}[b]{0.4875\textwidth}
    \includegraphics[width=\textwidth]{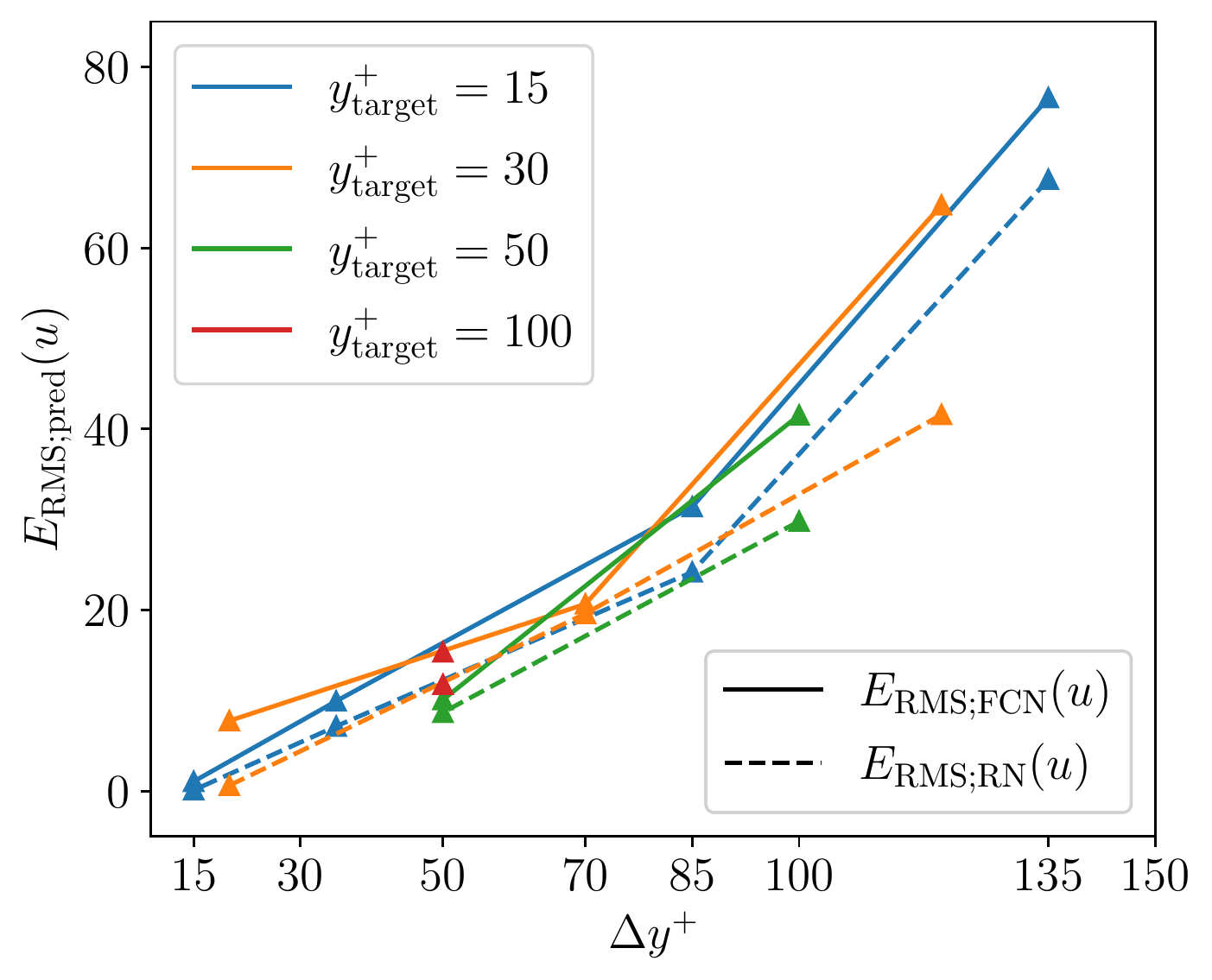}
  \end{minipage}
  \begin{center}
    \includegraphics[width=0.5\textwidth]{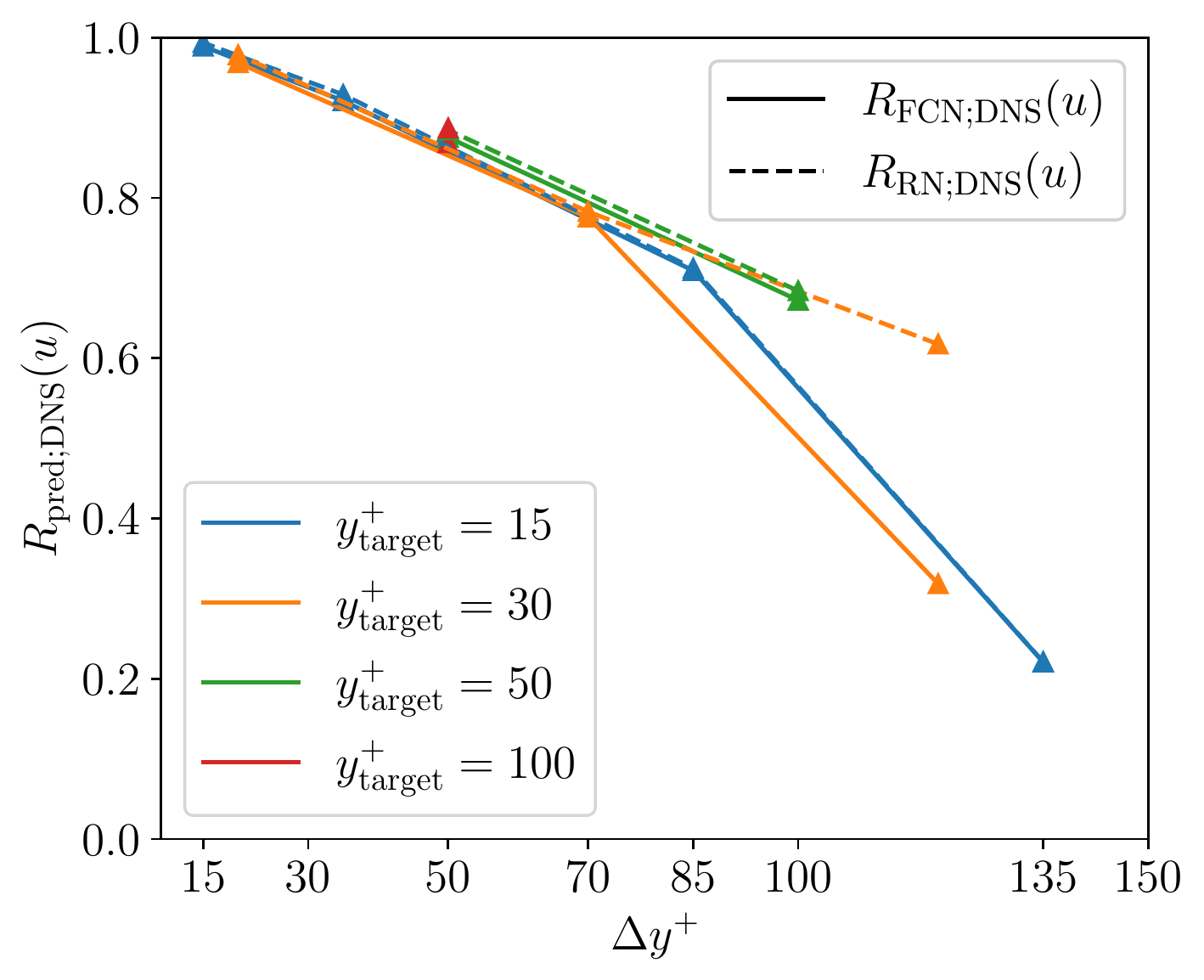}\\
    \end{center}
    \caption{\label{loss1}Variation of (top-left) mean-squared error normalized with the square of the RMS, (top-right) relative error in prediction of RMS fluctuation, (bottom) correlation coefficient between the predicted and DNS fields for streamwise velocity component with respect to separation distance for inner predictions.}
\end{figure}

The mean-squared error in the instantaneous predictions, the relative percentage error in the prediction of root-mean-squared (RMS) fluctuations and the correlation coefficient between the predicted fields from FCN and the DNS fields are shown in figure~\ref{loss1}, for the streamwise component of velocity fluctuations. The MSE grows almost linearly with respect to $\Delta y^{+}$. However, from a qualitative analysis of the predictions, the performance is significantly degraded for $\Delta y^{+} > 80$. 
This is perhaps more evident when the statistical error is considered. For larger separation distances, the error in predicting the streamwise fluctuation intensity is above 30\% and we also observe a difference in the slope of the curves as the separation distance increases for $\Delta y^+ > 80$.
The correlation coefficient between the predicted and the DNS fields shows a trend similar to that of the MSE. For smaller separation distances, the correlation coefficient is high and close to 1, then it rapidly decreases with $\Delta y^{+}$. From figure~\ref{loss1} we can conclude that a correlation coefficient $R_{\mathrm{FCN; DNS}}(u) \approx 0.8$ is the minimum requirement to have convincing predictions. Out of the three velocity components predicted by the FCN network, the streamwise result is used as a reference because it is the most energetic one, influencing most of the overall network performance. 
The other two velocity components predicted along with the streamwise one are typically of the similar magnitude or slightly worse in terms of both statistical error and correlation coefficient.


When the input and the target fields are separated by a large wall-normal distance ($\Delta y^{+} > 80$), it can be difficult for the networks to learn the relation between input and output. One possible explanation for this is the poor correlation between the input and the target fields for larger separation distances. For smaller $\Delta y^{+}$, there is a significant correlation between the flow features while for larger $\Delta y^{+}$ the correlation drops drastically and the quality of predictions is reduced. The study by~\cite{sasaki} showed the lack of coherence (analogous to correlation and defined in frequency domain) in the short wavelengths (\textit{i.e.} in the small scales) between the input and target velocity fields for large $\Delta y^{+}$. These observations indicate that, for larger $\Delta y^{+}$, it would be harder to predict the fields using FCN. Additionally, the higher errors in the prediction of the target fields with larger separation distances can be attributed to the limitation of the convolutional operation when it is used to predict the smaller scales in the target plane (which is closer to the wall). This is because the convolutional operation acts as a filter on the input data. Hence, it becomes difficult for the network to predict the high-frequency content in the output. 

If we consider the results at a given separation distance, the MSE in the streamwise component of velocity fluctuation increases as the predicted velocity field is closer to the wall. This is because the velocity-fluctuation fields closer to the wall are characterized by smaller scales and it is harder to predict them from the input velocity fields that may not clearly exhibit such behaviour. Additional complexity arises due to the reduced variability of features in the receptive field (the region of input field from which a single point in the target is obtained) of the input velocity planes, compared to the target velocity planes closer to the wall. We verified this hypothesis by inverting the input and target velocity fields (so that $y^{+}_{\rm input} < y^{+}_{\rm target}$) and training the network models to predict the flow farther away from the wall, using near-wall velocity fields as inputs. These predictions are named \textit{outer} predictions and they are similar to the ones performed by~\cite{guastoni_2022} except that the inputs are velocity-fluctuation fields at a given wall-normal distance, instead of the wall-shear stresses and the wall pressure. The inner predictions are more complicated because 
the smaller scales have to be inferred from the larger ones. Indeed, the network models provide a lower MSE in outer predictions compared to inner predictions due to the wide range of small scales in the input velocity fields closer to the wall. The results obtained for outer predictions are provided in~\ref{sec:outer_predictions}.

\subsection{Prediction in the self-similar region}
\label{subsec:self_similar_results}

\begin{figure}[H]
\begin{center}
\includegraphics*[width=1.0\linewidth]{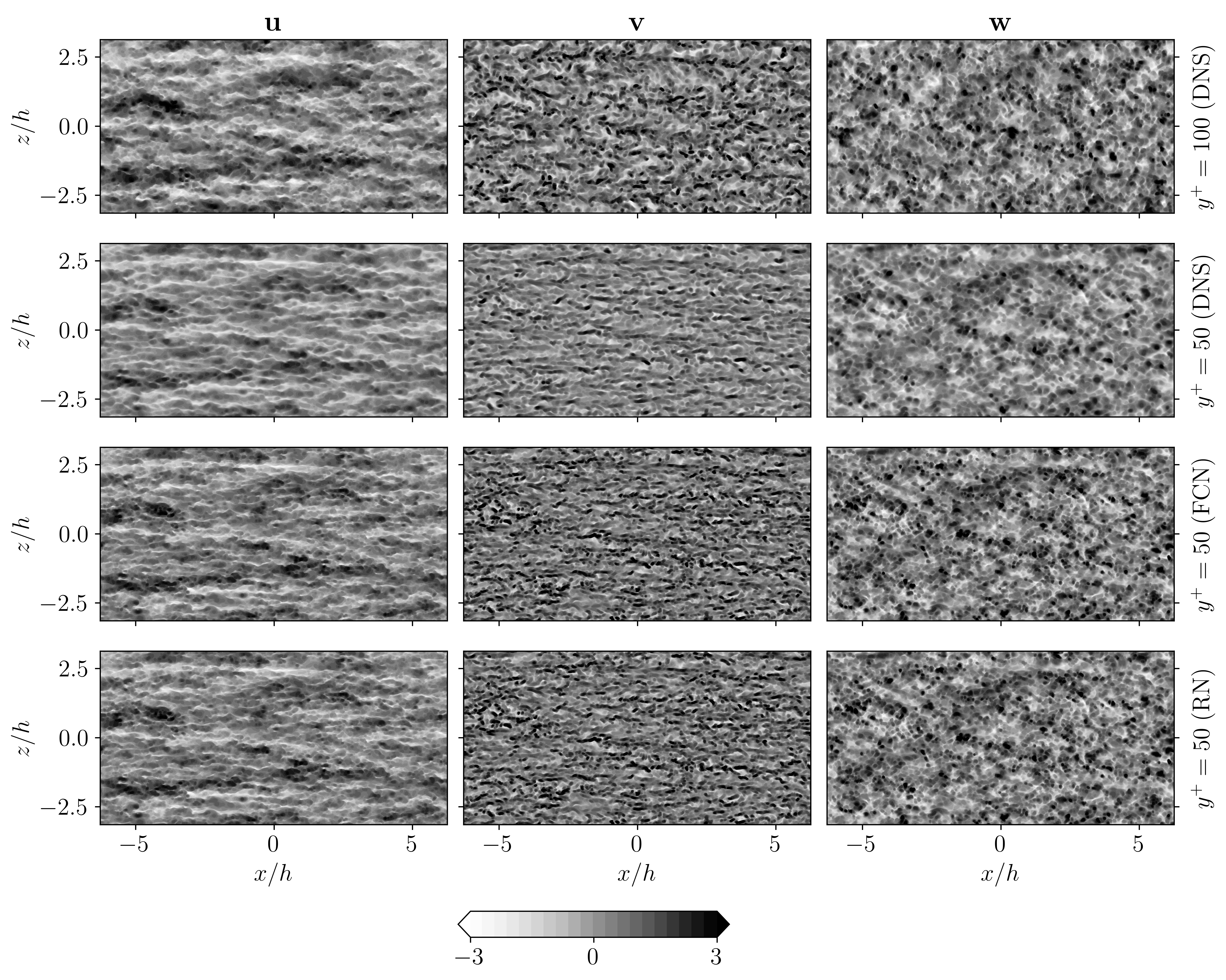}
\caption{\label{fcn_550} (Top row) DNS velocity fluctuations at $y^{+} = 100$, (second row) at $y^{+} = 50$ and (third row) corresponding FCN predictions, (fourth row) R-Net predictions of the (left column) streamwise, (middle column) wall-normal and (right column) spanwise velocity fluctuations at $y^{+}=50$ for $Re_{\tau} = 550$. The fields are scaled with respect to the corresponding RMS values.}   
\end{center}
\end{figure}

From the previous results, we find that it is challenging for the networks to find a non-linear transfer function between the input and output fields with larger $\Delta y^{+}$. At higher Reynolds number, despite the larger range of scales in the flow field, this task becomes simpler because it is possible to exploit the linear dependence of eddy size with respect to the distance from the wall within the logarithmic layer. To this end, we consider the predictions at $y^{+}=50$ $(y/h = 0.1)$ using the velocity-fluctuation fields at $y^{+} = 100$ $(y/h = 0.2)$ for $Re_{\tau} = 550$. These wall-normal locations are similar to the boundary and the reference planes considered by~\cite{mizuno} for the studied $Re_{\tau}$. It should be noted that the linear dependence was originally hypothesized at an asymptotic limit, at very high Reynolds number but, it was also used for finite Reynolds numbers. In figure~\ref{fcn_550} we show a qualitative comparison of the streamwise, wall-normal and spanwise velocity fluctuations predicted by the FCN and R-Net, compared with the reference DNS data. The quality of prediction is quantitatively assessed using the performance metrics introduced in~$\S$\ref{subsec:performance} and they are summarized in the tables~\ref{ret550_result_FCN},~\ref{ret550_result_UN} for the FCN and R-Net predictions at $y^{+} = 50$. Since the training procedure is stochastic, the reported errors in the predictions are averaged over 3 different models obtained with different initializations of the learnable parameters of the network.

\begin{table}[H]
\begin{center}
\caption{\label{ret550_result_FCN} Model-averaged errors in the FCN predictions of velocity fluctuation fields at $y^{+} = 50$ from $y^{+} = 100$ corresponding to $Re_{\tau} = 550$}
\begin{tabular}{cccc}
& & \multicolumn{1}{c}{${i}$} & \\ \cline{2-4}
Parameters           & ${u}$                    & ${v}$                      &${w}$\\
\hline
$\mathcal{L}({i}_\mathrm{FCN}; {i}_\mathrm{DNS})/i_\mathrm{RMS}^2$ & 0.241 $\pm$ 0.004 & 0.425 $\pm$ 0.009 & 0.324 $\pm$ 0.007 \\
$E_{\mathrm{RMS; FCN}} ({i}) \hspace{0.1cm} [\%] $ & 12.15 $\pm$ 1.33 & 22.1 $\pm$ 3.46 & 16.91 $\pm$ 2.19 \\
$R_{\mathrm{FCN;DNS}}$ & 0.871 $\pm$ 0.003 & 0.759 $\pm$ 0.006 & 0.821 $\pm$ 0.004 \\
\end{tabular}
\end{center}
\end{table}

 \begin{table}[H]
\begin{center}
\caption{\label{ret550_result_UN} Model-averaged errors in the R-Net predictions of velocity fluctuation fields at $y^{+} = 50$ from $y^{+} = 100$ corresponding to $Re_{\tau} = 550$}
\begin{tabular}{cccc}
& & \multicolumn{1}{c}{${i}$} & \\ \cline{2-4}
Parameters           & ${u}$                    & ${v}$                      &${w}$\\
\hline
$\mathcal{L}({i}_\mathrm{RN}; {i}_\mathrm{DNS})/i_\mathrm{RMS}^2$ & 0.239 $\pm$ 0.003 & 0.425 $\pm$ 0.004 & 0.322 $\pm$ 0.003 \\
$E_{\mathrm{RMS; RN}} ({i}) \hspace{0.1cm} [\%] $ & 10.81 $\pm$ 1.14 & 18.4 $\pm$ 1.22 & 14.58 $\pm$ 2.44 \\
$R_{\mathrm{RN;DNS}}$ & 0.873 $\pm$ 0.002 & 0.763 $\pm$ 0.002 & 0.824 $\pm$ 0.002 \\
\end{tabular}
\end{center}
\end{table}

\begin{figure}[H]
    \begin{center}
    \includegraphics[width=1.0\textwidth]{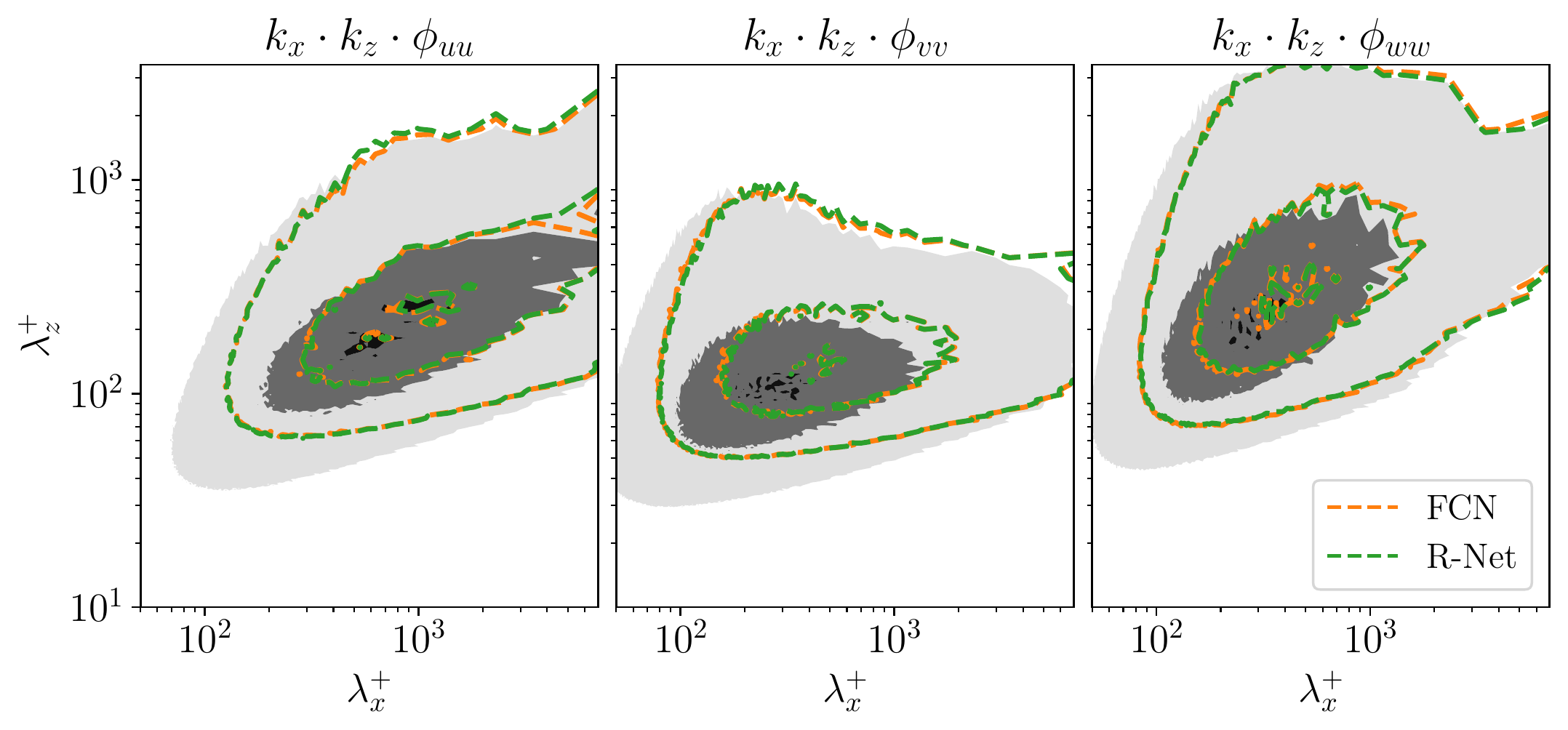}
    \end{center}
    \caption{\label{spectra} Pre-multiplied two-dimensional power-spectral densities for $Re_{\tau} = 550$ at $y^{+} = 50$. The contour levels contain 10\%, 50\% and 90\% of the maximum DNS power-spectral density. Shaded contours refer to DNS data and the dashed contour lines correspond to FCN and R-Net predictions.}
\end{figure}

At $y^{+} = 50$, the relative error in the RMS streamwise velocity fluctuations from the FCN and R-Net is around 10\%, which is similar to that observed in the low~$Re_\tau$ case. Also, the correlation between the predicted and the DNS velocity-fluctuation field in the streamwise direction is more than 85\% as can also be observed in tables~\ref{ret550_result_FCN} and~\ref{ret550_result_UN}. Note that the convolutional networks are not explicitly optimized to reproduce the statistical behaviour of the flow, so the results in this metric depend entirely on the capability of the neural network to predict the instantaneous velocity-fluctuation fields. From figure~\ref{fcn_550}, it can be observed that the large scales in the streamwise velocity-fluctuation fields are well represented in the predictions from FCN and R-Net. The predictions in the smaller scales are also in good agreement with the DNS reference. On the other hand, this is not true for the predictions of the wall-normal and the spanwise velocity-fluctuation fields, which appear smoother than their DNS counterpart, as if a low-pass filter was used. This qualitative analysis is confirmed by the pre-multiplied two-dimensional power-spectral density of the streamwise, wall-normal and spanwise fluctuations shown in figure~\ref{spectra}. From the spectra, it is possible to observe how the energy content of the velocity fields is more accurately reproduced in the streamwise direction, although all three velocity-component predictions lack energy at the shortest wavelengths.

This result highlights that the self-similarity can be implicitly utilized by the FCN in predicting the velocity-fluctuation fields in the overlap region at higher $Re_{\tau}$. 
These preliminary results indicate the advantage of incorporating the physical knowledge available for wall-bounded flows during the development of prospective data-driven wall models for LES. 

\subsection{Comparison of network performances}
\label{subsec:network_performance}
\begin{figure}[H]
    \begin{center}
    \includegraphics[width=1.0\textwidth]{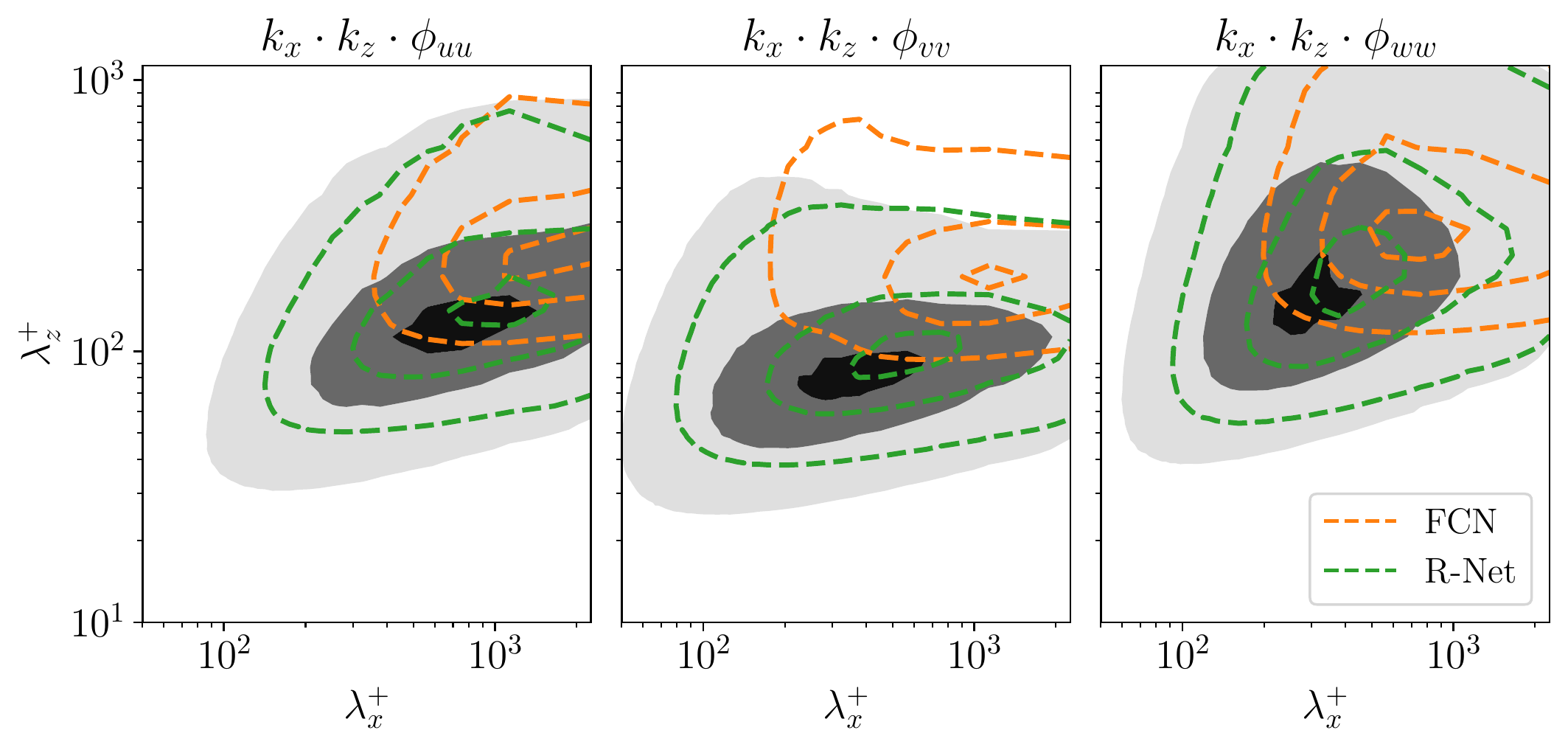}
    \end{center}
    \caption{\label{fig:spectra_network} Pre-multiplied two-dimensional power-spectral densities for $Re_{\tau} = 180$ at $y^{+} = 30$. The contour levels contain 10\%, 50\% and 90\% of the maximum DNS power-spectral density. Shaded contours refer to DNS data and the dashed contour lines correspond to FCN and R-Net predictions.}
\end{figure}

A brief comparison of the network performance between the FCN and the proposed R-Net architecture is performed, with a motivation towards identifying the capability of the network architectures in predicting the flow quantities closer to the wall and thereby enabling the choice of employing an efficient neural network in predicting wall quantities. Overall, from the results discussed in~$\S$\ref{subsec:Inner_predictions}--\ref{subsec:self_similar_results}, it is observed that both the network architectures are able to predict the velocity fluctuation fields closer to the wall with similar order of accuracy. However, from figure~\ref{loss1}, it is observed that for larger separation distances between the inputs and outputs, R-Net architecture tends to provide better quality of predictions in comparison to the results obtained from FCN. Whereas for smaller separation distances, the performance is very similar and in some cases slightly  better when compared with the performance of FCN. Further, the R-Net architecture is well-suited to capture the overall turbulent kinetic energy in the predictions as close as possible to the DNS when compared against the results from FCN. This is also observed from the comparison of the pre-multiplied power sepctral density plots obtained for predicting the velocity fluctuation fields at $y^+ = 30$ from the velocity fluctuation fields at $y^+ = 150$ for $Re_\tau = 180$ as shown in figure~\ref{fig:spectra_network}.

Although neither of the two architectures is able to capture the small scales in the velocity-fluctuation fields closer to the wall, the spectra of the predicted fluctuation fields from R-Net exhibits better agreement with that obtained from DNS. This indicates that the R-Net architecture is able to take advantage of the skip-connections in steering the network towards identifying the most energetic scales in the outputs. Further, the possible similarity in the large scales between the input and output fields are better exploited by the skip-connections in R-Net in comparison to the FCN architecture and hence, the R-Net architecture is able to predict well the most energetic scales in all components of the predictions.

In addition, it should also be noted that the size of the proposed R-Net architecture is about 11\% smaller than that of the FCN, which also highlights that the R-Net is able to reuse the features learnt in the upstream part of the network towards identifying the outputs with similar or better accuracy in comparison to FCN. Further, from tables~\ref{ret550_result_FCN} and~\ref{ret550_result_UN}, a similar magnitude of accuracy is obtained in predicting the fields in the self-similar region. Possibly, for larger separation distances we might also observe better predictions in R-Net relative to FCN for the fields corresponding to $Re_\tau = 550$, which was not verified in this study.

\subsection{Prediction of wall quantities}
\label{subsec:wall_predictions}

\begin{figure}[H]
    \begin{center}
    \includegraphics[width=1.0\textwidth]{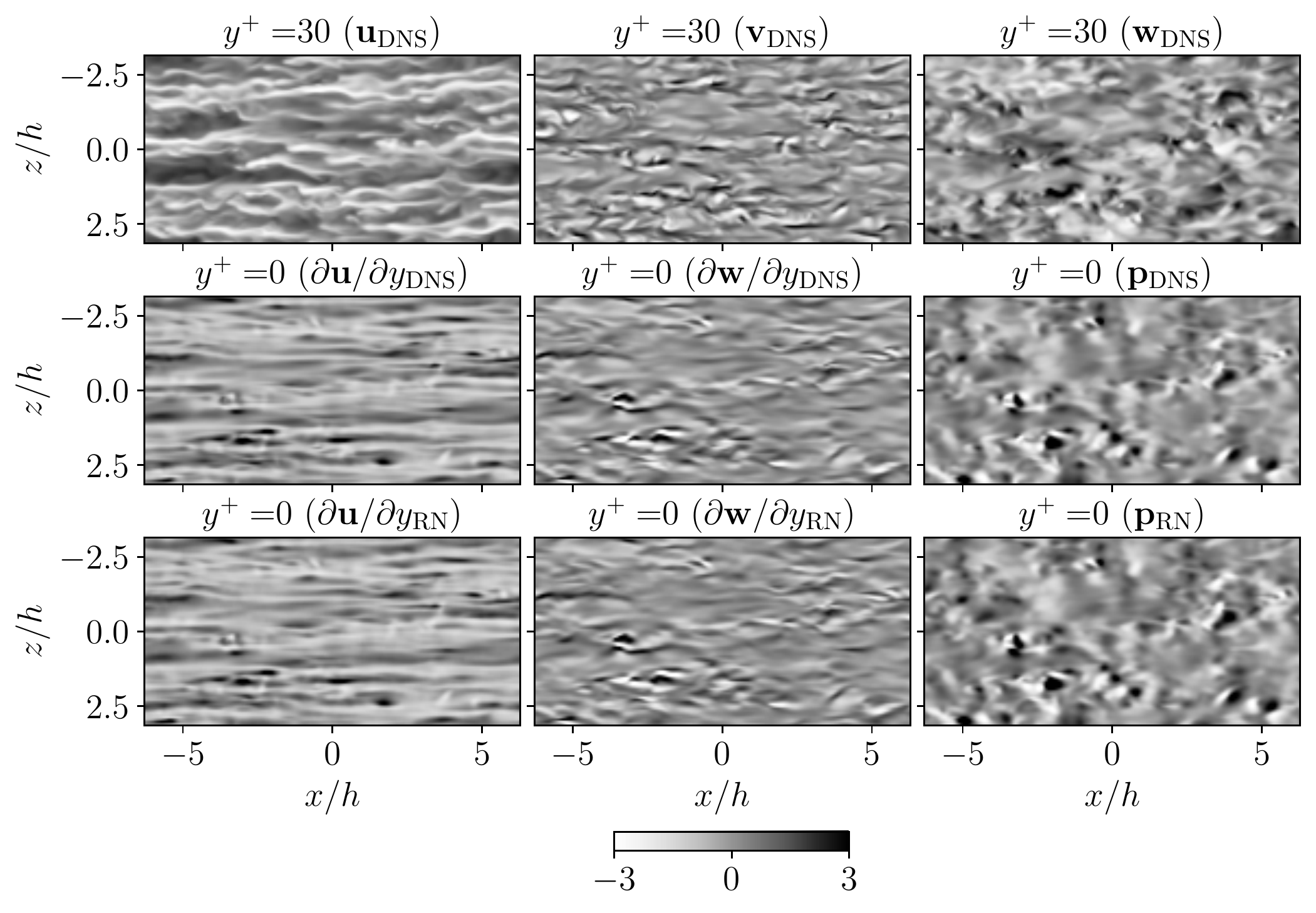}
    \end{center}
    \caption{\label{fig:ret180_wall_qual} (Top row) DNS velocity fluctuations at $y^{+} = 30$, (middle row) wall-quantities and (bottom row) corresponding R-Net predictions of the (left column) streamwise, (middle column) spanwise wall-shear and (right column) wall pressure for $Re_{\tau} = 180$. The fields are scaled with respect to the corresponding RMS values.}   
\end{figure}

\begin{figure}[H]
    \begin{center}
    \includegraphics[width=1.0\textwidth]{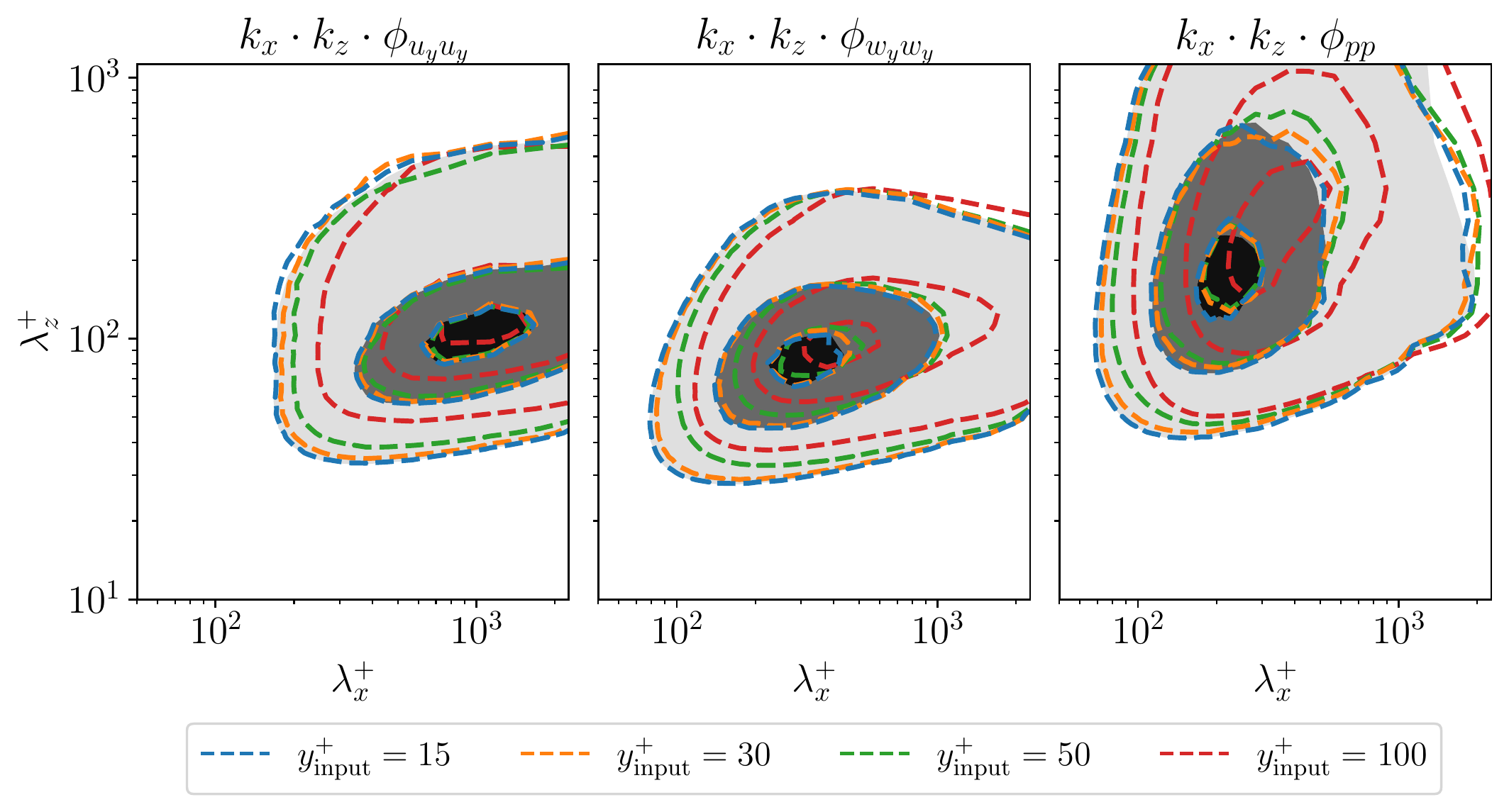}
    \end{center}
    \caption{\label{spectra_wall_ret180} Pre-multiplied two-dimensional power-spectral densities for the wall quantities at $Re_{\tau} = 180$. The contour levels contain 10\%, 50\% and 90\% of the maximum DNS power-spectral density. Shaded contours refer to DNS data and the dashed contour lines correspond to R-Net predictions with inputs at different wall-normal locations differentiated by the color of contour lines.}
\end{figure}

The proposed R-Net architecture is employed to predict the wall quantities such as streamwise wall-shear stress, spanwise wall-shear stress and wall pressure using the velocity-fluctuation fields at a certain $y^+$ plane as inputs. A sample qualitative prediction of the wall quantities corresponding to $Re_\tau = 180$, obtained using velocity-fluctuation fields at $y^+=30$ as inputs is shown in figure~\ref{fig:ret180_wall_qual}. Further, the network performance measures as introduced in~$\S$\ref{subsec:performance} is reported in table~\ref{ret180_result_UN_wall} for the different wall predictions conducted at $Re_\tau = 180$. The R-Net architecture is able to predict the wall-shear and wall pressure using the velocity fluctuation fields at $y^+ = 50$ with less than about 10\% error in the prediction of corresponding RMS quantities. Further, as the inner-scaled separation distance becomes larger than 80, (\textit{i.e.} in this case predicting the wall quantities using the fields at $y^+ = 100$,) we find the quality of predictions to drastically diminish. On the other hand, if the input fields are close to the wall (\textit{i.e.} $y^+\le 30$,) the prediction of wall quantities is in good agreement with the DNS fields and the error in the corresponding RMS quantities is less than about 2\% in almost all the components. A comparison of the premultiplied two-dimensional power spectral density between the DNS fields and the predictions is provided in figure~\ref{spectra_wall_ret180}. From the figure, it is observed that as the position of the input fields moves away from the wall, the network model lacks the capability to predict the small scale structures in the wall fields, which is closely related to the reasons outlined in~$\S$\ref{subsec:Inner_predictions}.

\begin{figure}[H]
    \begin{center}
    \includegraphics[width=1.0\textwidth]{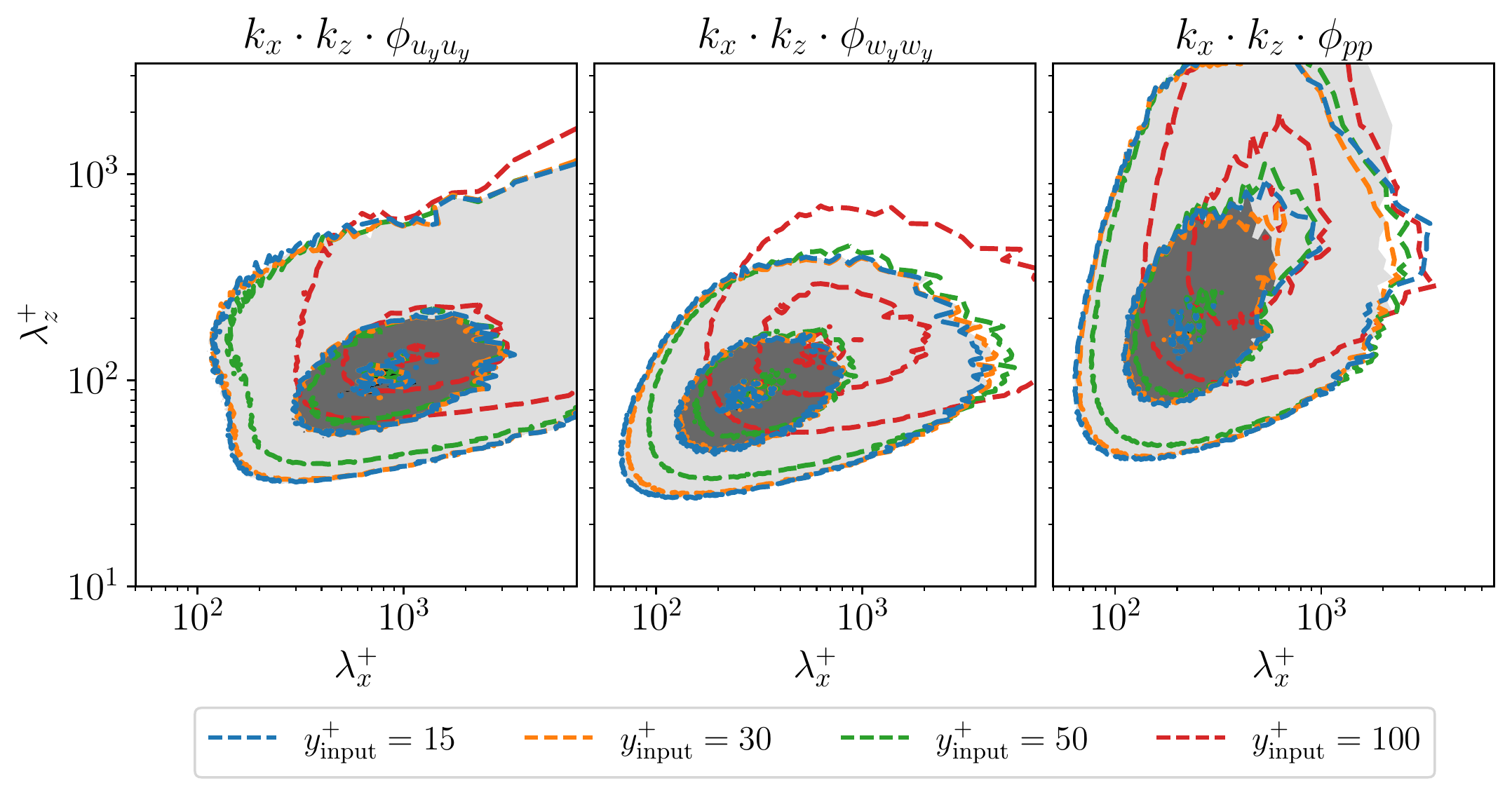}
    \end{center}
    \caption{\label{spectra_wall_ret550} Pre-multiplied two-dimensional power-spectral densities for the wall quantities at $Re_{\tau} = 550$. The contour levels contain 10\%, 50\% and 90\% of the maximum DNS power-spectral density. Shaded contours refer to DNS data and the dashed contour lines correspond to R-Net predictions with inputs at different wall-normal locations differentiated by the color of contour lines.}
\end{figure}

Similarly, for the wall predictions at $Re_\tau = 550$, the error metrics are provided in table~\ref{ret550_result_UN_wall}. A sample qualitative prediction of the wall quantities corresponding to $Re_\tau = 550$, obtained using velocity-fluctuation fields at $y^+=50$ is also provided in figure~\ref{fig:ret180_wall_qual}. Finally, a comparison of the premultiplied two-dimensional power spectral density between the DNS fields and the predictions for $Re_\tau=550$ is provided in figure~\ref{spectra_wall_ret550}. The observations made for the wall predictions at $Re_\tau = 180$ is readily applicable to the predictions at $Re_\tau = 550$. Further, the magnitude of error is not drastically different from that observed at $Re_\tau = 180$, although the errors at $Re_\tau = 550$ are higher due to the wider range of scales contained in the fields to be predicted.

From the results discussed above, we observe the prediction quality to drastically improve as the input fields are closer to the wall. Hence for predicting the wall quantities at $Re_\tau = 550$ from the input fields at $y^+ = 100$, it is intuitive to take advantage of the self-similarity and implement a convolutional network to predict the fields at $y^+ = 50$ and thereby to utilize a R-Net architecture to finally obtain the wall quantities. This approach utilizes the velocity fluctuations at $y^+=50$ as auxillary quantities in the prediction of wall-shear and wall-pressure. However, such an approach also yielded results which were similar to the ones observed in table~\ref{ret550_result_UN_wall}. 

 \begin{table}[H]
\begin{center}
\caption{\label{ret180_result_UN_wall} Model-averaged errors in the R-Net predictions of wall-quantities corresponding to $Re_{\tau} = 180$}
\begin{tabular}{ccccc}
& & & \multicolumn{1}{c}{${i}$} & \\ \cline{3-5}
\multicolumn{1}{c|}{$y^+_{\rm inputs}$} & Parameters           & ${u_y}$                    & ${w_y}$                      &${p}$\\
\hline
\multicolumn{1}{c|}{}&  $\mathcal{L}({i}_\mathrm{RN}; {i}_\mathrm{DNS})/i_\mathrm{RMS}^2$ & 0.012 $\pm$ 0.002 & 0.022 $\pm$ 0.006 & 0.099 $\pm$ 0.008 \\
\multicolumn{1}{c|}{15}& $E_{\mathrm{RMS; RN}} ({i}) \hspace{0.1cm} [\%] $ & 1.60 $\pm$ 1.09 & 2.97 $\pm$ 0.95 & 2.09 $\pm$ 0.86 \\
\multicolumn{1}{c|}{}& $R_{\mathrm{RN;DNS}}$ & 0.994 $\pm$ 0.001 & 0.993 $\pm$ 0.001 & 0.949 $\pm$ 0.004 \\
\hline
\multicolumn{1}{c|}{}&  $\mathcal{L}({i}_\mathrm{RN}; {i}_\mathrm{DNS})/i_\mathrm{RMS}^2$ & 0.056 $\pm$ 0.001 & 0.070 $\pm$ 0.003 & 0.107 $\pm$ 0.006 \\
\multicolumn{1}{c|}{30}& $E_{\mathrm{RMS; RN}} ({i}) \hspace{0.1cm} [\%] $ & 2.62 $\pm$ 1.28 & 2.34 $\pm$ 0.92 & 3.07 $\pm$ 2.21 \\
\multicolumn{1}{c|}{}& $R_{\mathrm{RN;DNS}}$ & 0.971 $\pm$ 0.001 & 0.965 $\pm$ 0.001 & 0.945 $\pm$ 0.003 \\ 
\hline
\multicolumn{1}{c|}{}&  $\mathcal{L}({i}_\mathrm{RN}; {i}_\mathrm{DNS})/i_\mathrm{RMS}^2$ & 0.244 $\pm$ 0.013 & 0.349 $\pm$ 0.009  & 0.272 $\pm$ 0.010 \\
\multicolumn{1}{c|}{50}& $E_{\mathrm{RMS; RN}} ({i}) \hspace{0.1cm} [\%] $ & 10.31 $\pm$ 3.10 & 14.73 $\pm$ 3.30 & 13.07 $\pm$ 3.06 \\
\multicolumn{1}{c|}{}& $R_{\mathrm{RN;DNS}}$ & 0.848 $\pm$ 0.004 & 0.815 $\pm$ 0.004 & 0.853 $\pm$ 0.006 \\ 
\hline
\multicolumn{1}{c|}{}&  $\mathcal{L}({i}_\mathrm{RN}; {i}_\mathrm{DNS})/i_\mathrm{RMS}^2$ & 0.674 $\pm$ 0.003 & 0.709 $\pm$ 0.002 & 0.620 $\pm$ 0.006 \\
\multicolumn{1}{c|}{100}& $E_{\mathrm{RMS; RN}} ({i}) \hspace{0.1cm} [\%] $ & 33.31 $\pm$ 0.84 & 43.77 $\pm$ 0.86 & 39.26 $\pm$ 2.48 \\
\multicolumn{1}{c|}{}& $R_{\mathrm{RN;DNS}}$ & 0.574 $\pm$ 0.002 & 0.538 $\pm$ 0.001 & 0.616 $\pm$ 0.006 

\end{tabular}
\end{center}
\end{table}

\begin{table}[H]
\begin{center}
\caption{\label{ret550_result_UN_wall} Model-averaged errors in the R-Net predictions of wall-quantities corresponding to $Re_{\tau} = 550$}
\begin{tabular}{ccccc}
& & & \multicolumn{1}{c}{${i}$} & \\ \cline{3-5}
\multicolumn{1}{c|}{$y^+_{\rm inputs}$} & Parameters           & ${u_y}$                    & ${w_y}$                      &${p}$\\
\hline
\multicolumn{1}{c|}{}&  $\mathcal{L}({i}_\mathrm{RN}; {i}_\mathrm{DNS})/i_\mathrm{RMS}^2$ & 0.013 $\pm$ 0.0003 & 0.016 $\pm$ 0.0013 & 0.137 $\pm$ 0.013 \\
\multicolumn{1}{c|}{15}& $E_{\mathrm{RMS; RN}} ({i}) \hspace{0.1cm} [\%] $ & 0.89  $\pm$ 0.68 & 1.51 $\pm$ 1.12 & 3.07 $\pm$ 1.97 \\
\multicolumn{1}{c|}{}& $R_{\mathrm{RN;DNS}}$ & 0.993 $\pm$ 0.0002 &  0.992 $\pm$ 0.0003 & 0.922 $\pm$ 0.0068 \\
\hline
\multicolumn{1}{c|}{}&  $\mathcal{L}({i}_\mathrm{RN}; {i}_\mathrm{DNS})/i_\mathrm{RMS}^2$ & 0.074 $\pm$ 0.0001 & 0.088 $\pm$ 0.0032 & 0.133 $\pm$ 0.0002 \\
\multicolumn{1}{c|}{30}& $E_{\mathrm{RMS; RN}} ({i}) \hspace{0.1cm} [\%] $ & 3.90 $\pm$ 0.44 & 2.91 $\pm$ 2.24 & 2.92 $\pm$ 2.23 \\
\multicolumn{1}{c|}{}& $R_{\mathrm{RN;DNS}}$ & 0.962 $\pm$ 0.0001 & 0.956 $\pm$ 0.0002 & 0.931 $\pm$ 0.0008\\ 
\hline
\multicolumn{1}{c|}{}&  $\mathcal{L}({i}_\mathrm{RN}; {i}_\mathrm{DNS})/i_\mathrm{RMS}^2$ & 0.257 $\pm$ 0.0031 & 0.316  $\pm$ 0.0045 & 0.228 $\pm$ 0.0042 \\
\multicolumn{1}{c|}{50}& $E_{\mathrm{RMS; RN}} ({i}) \hspace{0.1cm} [\%] $ & 10.35 $\pm$ 1.53 & 13.07 $\pm$ 1.62 & 8.68 $\pm$ 0.24 \\
\multicolumn{1}{c|}{}& $R_{\mathrm{RN;DNS}}$ & 0.865 $\pm$ 0.0020 & 0.827 $\pm$ 0.0032 & 0.879 $\pm$ 0.0023 \\ 
\hline
\multicolumn{1}{c|}{}&  $\mathcal{L}({i}_\mathrm{RN}; {i}_\mathrm{DNS})/i_\mathrm{RMS}^2$ & 0.837 $\pm$ 0.0035 & 0.906 $\pm$ 0.0202 & 0.736 $\pm$ 0.0893 \\
\multicolumn{1}{c|}{100}& $E_{\mathrm{RMS; RN}} ({i}) \hspace{0.1cm} [\%] $ & 43.86 $\pm$ 3.17 & 58.08 $\pm$ 4.77 & 25.07 $\pm$ 9.53 \\
\multicolumn{1}{c|}{}& $R_{\mathrm{RN;DNS}}$ & 0.431 $\pm$ 0.0112 & 0.323 $\pm$ 0.0321 & 0.558 $\pm$ 0.0411

\end{tabular}
\end{center}
\end{table}

\begin{figure}[H]
    \begin{center}
    \includegraphics[width=1.0\textwidth]{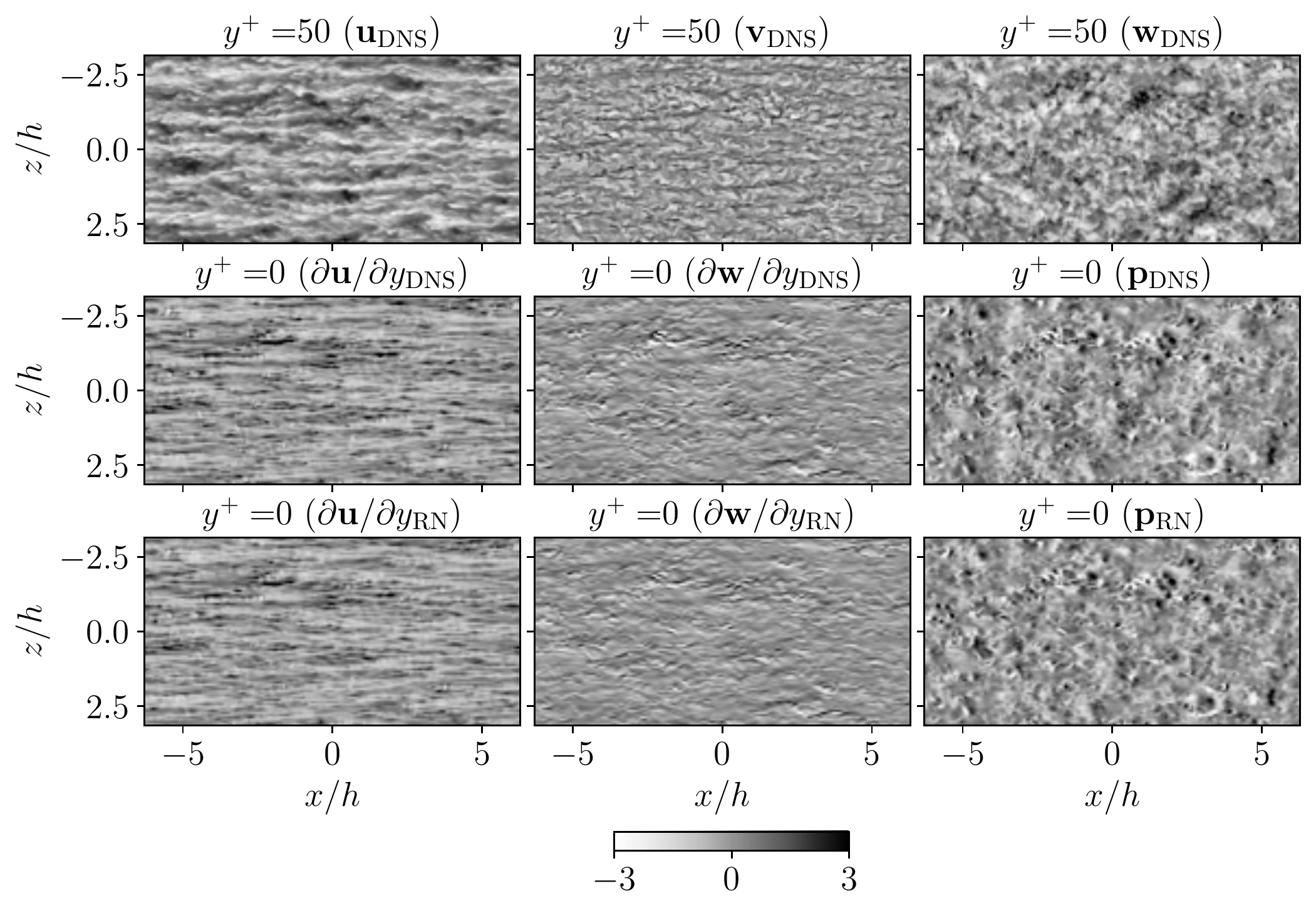}
    \end{center}
    \caption{\label{fig:ret550_wall_qual} (Top row) DNS velocity fluctuations at $y^{+} = 50$, (middle row) wall-quantities and (bottom row) corresponding R-Net predictions of the (left column) streamwise, (middle column) spanwise wall-shear and (right column) wall pressure for $Re_{\tau} = 550$. The fields are scaled with respect to the corresponding RMS values.}   
\end{figure}

ter\section{Conclusions}
\label{sec:conclusion}
In this work, the possibility of predicting the velocity-fluctuation fields closer to the wall using the fluctuation fields farther from it by means of an FCN and R-Net is assessed. Several predictions were performed to understand the implementation and the quality of predictions at lower $Re_{\tau}$. The variation of the MSE with respect to the separation distance between the input and the target fluctuation fields exhibits a non-linear behaviour and it also varies with respect to the $y^{+}$ location of the predicted fluctuation field. The results also indicate the capability of the FCN to predict the non-linear transfer function between the velocity-fluctuation fields that are separated by short distances.

Additionally, a higher $Re_{\tau}$ of 550 was also considered for testing the capability of the network to predict fields closer to the wall characterized by  wider range of scales. At high Reynolds number, the self-similarity hypothesis in the logarithmic layer can be exploited by the convolutional networks. The R-Net architecture is used to predict the velocity-fluctuation fields at $y^{+}=50$ using the fluctuation fields at $y^{+}=100$ corresponding to $Re_{\tau} = 550$ and it provides better instantaneous predictions than the lower-Reynolds case of $Re_{\tau}=180$ due to the corresponding velocity-fluctuation fields at $y/h \approx 0.27$ and $0.55$, which are not in self-similar region. The fluctuation intensities are also reasonably well-predicted for $Re_{\tau}=550$, with an error in the streamwise RMS-fluctuation is around 10\% compared to the DNS statistics. Furthermore, the spectral analysis of the predictions shows that the energy content at the length-scales that are present in the input flow field are well-predicted. Note, however, that the energy in the smaller scales is difficult to reproduce since the input fields do not provide information about them.

The present results highlight that the self-similarity in the overlap region of the flow can be effectively utilized by the network models to predict the velocity-fluctuation fields at higher $Re_{\tau}$. Note that such similarity is not explicitly enforced in the neural-network architecture, but this physical property can be exploited by the network to obtain better predictions. Further, the study is extended to predict the wall-shear stress components and wall pressure using the velocity-fluctuation fields at a certain wall-normal distance close to the wall. The proposed R-Net architecture is able to predict the wall quantities with less than 15\% error in the corresponding fluctuation intensity at both $Re_\tau = 180$ and $550$, when the input velocity fluctuation fields at $y^+ = 50$ are used. The closer to the wall the input fields are, the higher is the prediction accuracy. These results are an encouraging starting point to utilize such neural-network models in the development of data-driven wall models for LES. In the direction of improving accuracy of predictions obtained from network models, the use of U-Net architecture to reconstruct the small features in the wall is a viable option and will be considered in the scope of future work. Note however that this is just a proof of concept regarding the predictive capabilities of these models, and being able to implement such strategies in \textit{a-posteriori} tests would require combining these methods with a LES solver.

\section{Acknowledgments}
The authors acknowledge the funding provided by the Swedish e-Science Research Centre (SeRC), ERC grant no.~"2021-CoG-101043998, DEEPCONTROL" to RV and the Knut and Alice Wallenberg (KAW) Foundation. The authors thank Dr. Alejandro G\"uemes, Prof. Andrea Ianiro and Prof. Stefano Discetti for their valuable insights and discussions on this research project. The analysis was performed on resources provided by the Swedish National Infrastructure for Computing (SNIC) at PDC. Views and opinions expressed are however those of the author(s) only and do not necessarily reflect those of the European Union or the European Research Council. Neither the European Union nor the granting authority can be held responsible for them.


\appendix

\section{Outer predictions}
\label{sec:outer_predictions}
\begin{figure}[H]
    \begin{center}
  \begin{minipage}[b]{0.5\textwidth}
    \includegraphics[width=\textwidth]{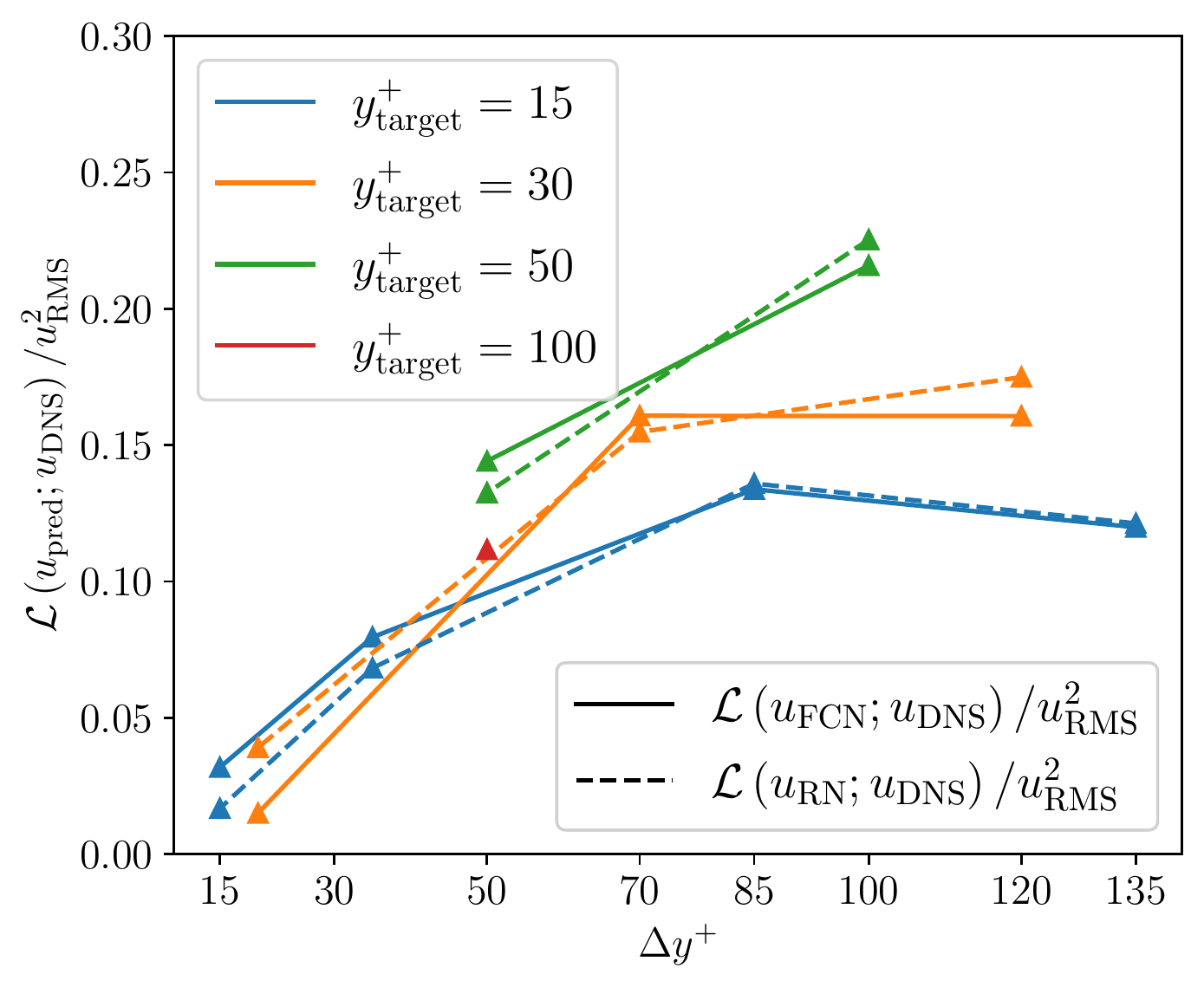}
  \end{minipage}
  \hfill
  \begin{minipage}[b]{0.485\textwidth}
    \includegraphics[width=\textwidth]{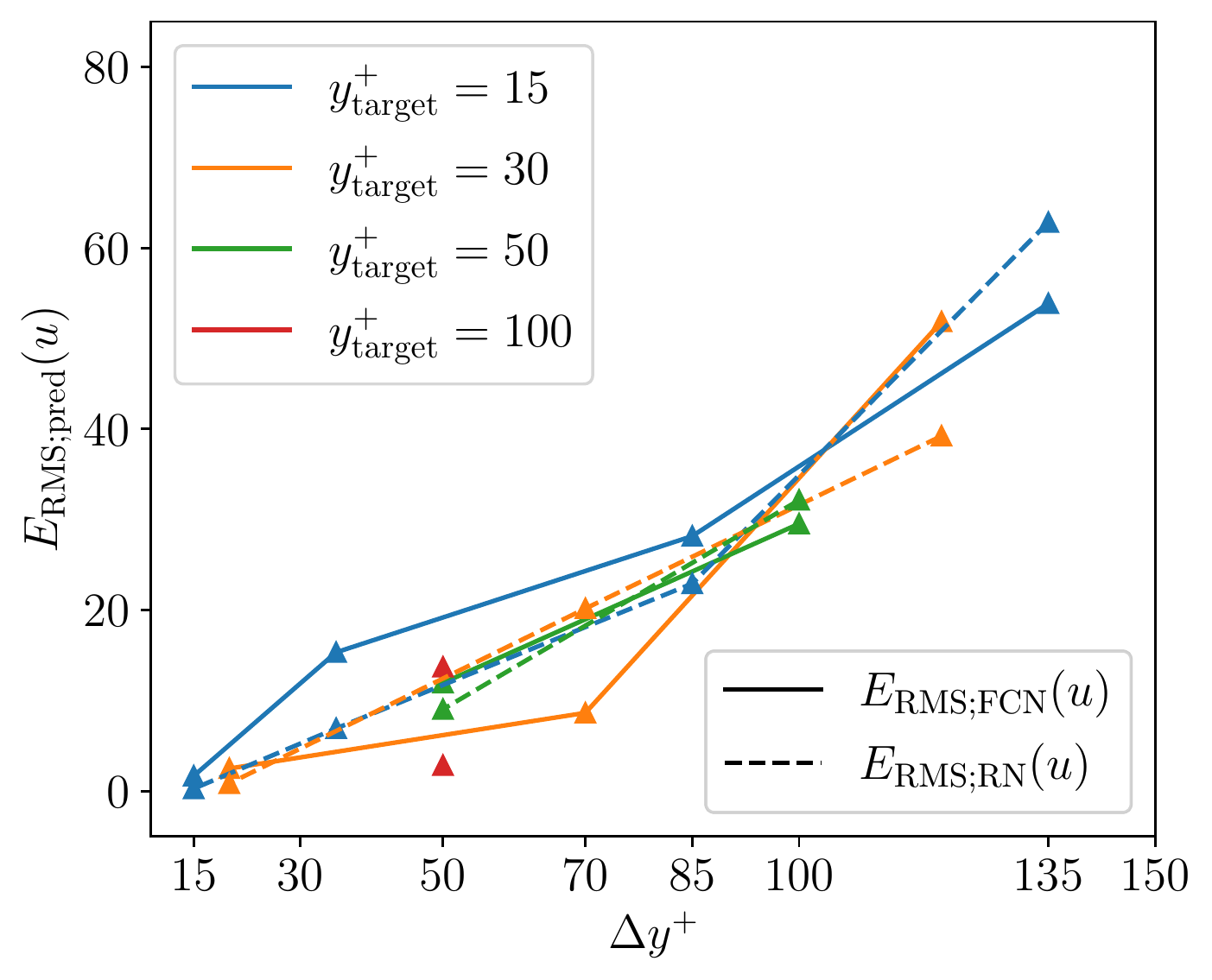}
  \end{minipage}
    \includegraphics[width=0.5\textwidth]{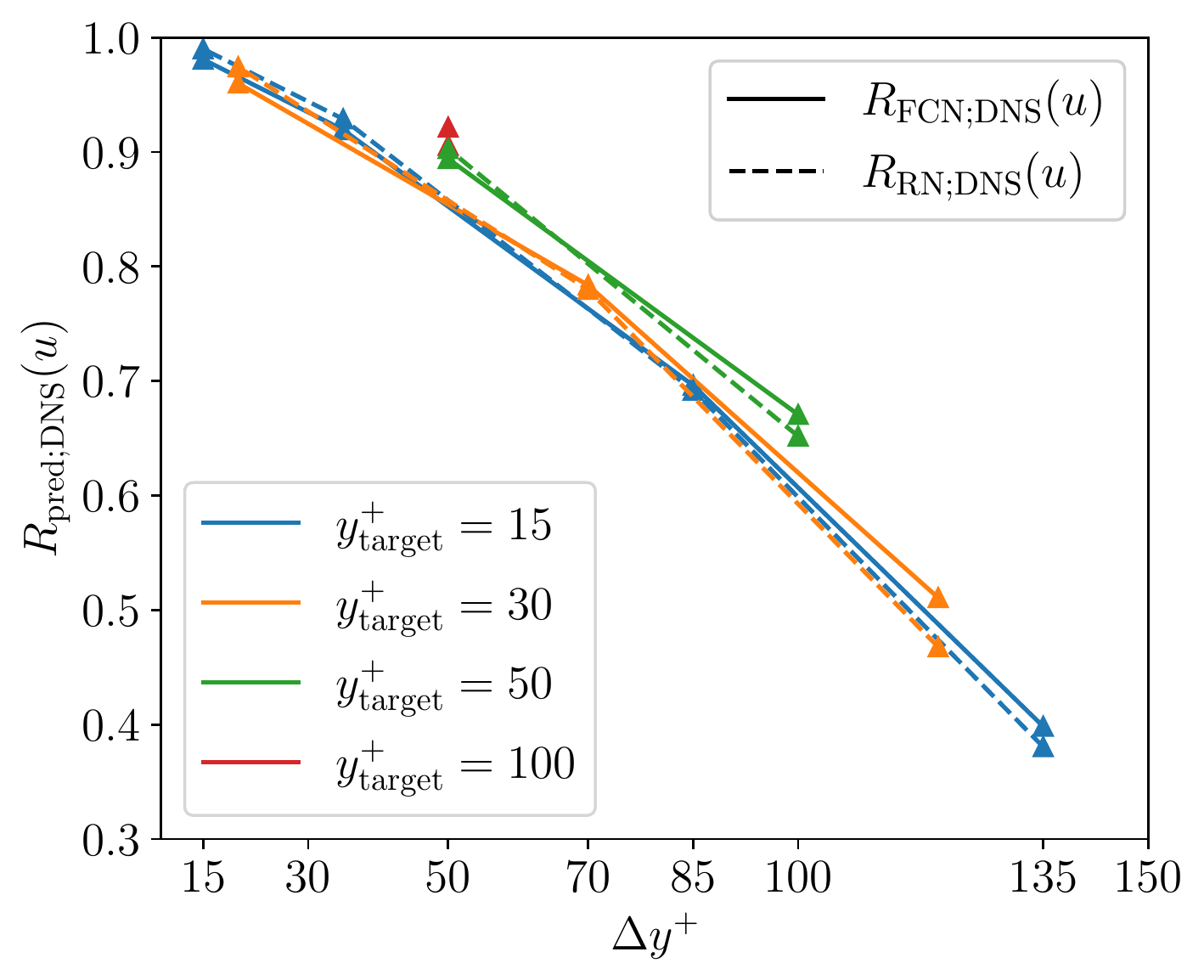}\\
    \end{center}
    \caption{\label{fig:mse_outer}Variation of (top-left) mean-squared error normalized with the square of the RMS, (top-right) relative error in prediction of RMS fluctuation, (bottom) correlation coefficient between the predicted and DNS fields for streamwise velocity component with respect to separation distance for outer predictions.}
\end{figure}

In contrast to the inner predictions discussed in~$\S$\ref{subsec:Inner_predictions}, outer predictions refer to the prediction of velocity fluctuation fields farther from the wall using the fields closer to the wall as inputs. A total of 10 predictions are performed with the velocity fields obtained at various wall-normal locations from the DNS simulation at $Re_\tau = 180$. Here, the separation distances are defined as: $\Delta y^{+} = y^{+}_{\rm target} - y^{+}_{\rm input}$. The input velocity-fluctuation field closest to the wall is located at $y^{+} = 15$ and the corresponding target field is at $y^{+} = 30$. The results obtained for outer predictions are plotted in figure~\ref{fig:mse_outer}. In comparison to the results obtained for inner predictions as plotted in figure~\ref{loss1}, the outer predictions are observed to be much better and exhibits a different slope for the variation of error in prediction of streamwise fluctuation intensity with respect to separation distance. Further, a saturation of MSE for higher separation distances is clearly observed for inner predictions whereas such saturation was not conceived in the present study for outer predictions. If a particular separation distance is considered, the target fields closer to the wall are more accurately predicted in the case of outer predictions, a result opposite to what is observed for the inner predictions.

\bibliographystyle{elsarticle-harv} 
\bibliography{cas-refs}





\end{document}